\documentclass[11pt,a4paper]{article}
 \pdfoutput=1
\usepackage{jheppub}
\usepackage[utf8]{inputenc}
\usepackage[english]{babel}
\usepackage{enumitem}
\usepackage{subdepth}
\newcommand{\mylabel}[2]{#2\def\@currentlabel{#2}\label{#1}}
\makeatletter

 \def\be{\begin{equation}}
\def\ee{\end{equation}}
 \def\ba{\begin{align}}
\def\ea{\end{align}}
\def\bea{\begin{eqnarray}}
\def\eea{\end{eqnarray}}

\def\m{\mu}

\def\D{\mathcal D}
\def\ds{d}

\def\Cur{{\cal R}}

\newcommand{\bseq}{\begin{subequations}}
\newcommand{\eseq}{\end{subequations}}

\renewcommand{\ln}{\mathop{\rm ln}\nolimits}

\newcommand{\Tr}{{\rm Tr}}
\newcommand{\tr}{{\rm tr}}

\newcommand{\di}{\mathrm d}

\renewcommand{\AA}{B}

\title{Heat kernel methods for Lifshitz theories}

\author[a,d]{Andrei O.~Barvinsky,}
\author[b]{Diego~Blas,}
\author[c]{Mario~Herrero-Valea,}
\author[a]{Dmitry V.~Nesterov,}
\author[e]{Guillem~P\'erez-Nadal,}
\author[f]{Christian~F.~Steinwachs}

\affiliation[a]{ \it Theory Department, Lebedev Physics Institute, Leninskii Pr. 53, Moscow 119991, Russia}
\affiliation[b]{\it Theoretical Physics Department, CERN, CH-1211 Geneva 23,
 Switzerland}
\affiliation[c]{Institute of Physics, Laboratory of Particle Physics and Cosmology,\\
Ecole Polytechnique F\'ed\'erale de Lausanne, CH-1015, Lausanne, Switzerland}
\affiliation[d]{\it Tomsk State University, Department of Physics, Lenin Ave. 36, Tomsk 634050, Russia}
\affiliation[e]{\it Departamento de F\'isica, FCEN, Universidad de Buenos Aires.\\
 Ciudad Universitaria, Pabell\'on 1, 1428 Buenos Aires, Argentina}
\affiliation[f]{\it Physikalisches Institut, Albert-Ludwigs-Universit\"at Freiburg,\\
Hermann-Herder-Strasse 3, 79104 Freiburg, Germany}

\emailAdd{barvin@td.lpi.ru}
\emailAdd{diego.blas@cern.ch}
\emailAdd{mario.herrerovalea@epfl.ch}
\emailAdd{nesterov@td.lpi.ru}
\emailAdd{guillem@df.uba.ar}
\emailAdd{christian.steinwachs@physik.uni-freiburg.de}

\abstract{We study the one-loop covariant effective action of Lifshitz theories using the heat kernel technique. The characteristic feature of Lifshitz theories is an anisotropic scaling between space and time. This is enforced by the existence of a preferred foliation of space-time, which breaks Lorentz invariance. In contrast to the relativistic case, covariant Lifshitz theories are only invariant under diffeomorphisms preserving the foliation structure. We develop a systematic method to reduce the calculation of the effective action for a generic Lifshitz operator to an algorithm acting on known results for relativistic operators. In addition, we present techniques that drastically simplify the calculation for operators with special properties.  We demonstrate the efficiency of these methods by explicit applications. }

\begin{document}
\preprint{CERN-TH-2017-055, FR-PHENO-2017-006}

\maketitle

\section{Introduction}\label{sec:intro}

Quantum field theory (QFT) has proven to be the correct language to describe many
phenomena of Nature. This is true for a variety of energy scales, groups of symmetry and even dimensionality.
A canonical example is the {\it standard model of particle physics}, formulated as a QFT based on unitarity, Lorentz invariance and a particular
structure of internal gauge symmetries. This structure is by no means necessary for the consistency of QFT and can be enlarged (e.g. by supersymmetry or conformal symmetry)
or reduced. In this work we will  be interested in the latter, in particular in cases where the boost transformations of the Lorentz group are no longer a symmetry of the system.
This is the natural situation in {\it condensed matter} physics, e.g  \cite{Fradkin:1991nr,Ardonne:2003wa}. It has also been suggested that this may happen at a fundamental level,
since a renormalizable quantum theory of gravity in 4-dimensions seems possible by imposing anisotropy between space and time, a defining feature of {\it Lifshitz theories}  \cite{Horava:2009uw,Barvinsky:2015kil}\footnote{Another interesting example where one may want to consider Lorentz breaking theories to get consistent QFT is massive gravity, cf. \cite{Blas:2014ira} and references therein.}.
Lifshitz theories will be the main focus of our work.

Despite the importance of Lifshitz theories, many of the standard results of Lorentz invariant theories have not yet been extended to them. An incomplete list of achievements includes studies on Goldstone modes \cite{Nicolis:2012vf,Watanabe:2014fva,Griffin:2015hxa}, OPE \cite{Goldberger:2014hca}, correlation functions \cite{Keranen:2016ija}, holographic dualities \cite{Griffin:2011xs,Griffin:2012qx,Kachru:2008yh,Nakayama:2012sn,
Roychowdhury:2015cva,Fadafan:2015iwa,Taylor:2015glc,Foster:2016abe}, anisotropic Weyl anomalies and Ward identities \cite{Adam:2009gq,Arav:2014goa,Gomes:2011di,Baggio:2011ha,Arav:2016xjc,
Pal:2016rpz,Auzzi:2016lxb,Pal:2017ntk}, effective actions of Lifshitz scalar theories  \cite{Nesterov:2010yi,D'Odorico:2015yaa}, renormalization of different proposals \cite{Anselmi:2007ri,Iengo:2009ix,Giribet:2010th,LopezNacir:2011mt,Barvinsky:2015kil,Griffin:2017wvh}, studies on entanglement entropy \cite{Nesterov:2010yi,Zhou:2016ekn,Parker:2017lnh}, gauging and coset construction \cite{Karananas:2016hrm} and formal aspects of anisotropic Weyl invariance \cite{Perez-Nadal:2016tzr}. In this work we provide general methods for the {\it heat kernel} approach to the {\it one-loop effective action} for Lifshitz theories which can be applied to particular physical models.

The {\it heat kernel} technique is a powerful tool in quantum field theory to  calculate the quantum effective action in an arbitrary field background -- a fundamental quantity which contains almost all the information about the quantum system and from which many of the results of quantization can be derived directly. This approach, initially originating from asymptotic expansion methods for solutions of partial differential equations, turned out to be very efficient both in physical and mathematical applications \cite{DeWittbook:1965,McKean-Singer,Gilkeybook:1984} (see also \cite{Barvinsky-Scholarpedia}).  Indeed, it became an indispensable tool for the analysis of renormalizability properties in generic models of gauge theories and quantum gravity. Thus far, this method was mainly
developed within generally covariant (and, therefore, Lorentz-invariant) theories with a
special emphasis
on the class of the so-called {\it minimal} second-order operators \cite{Barvinsky:1985an,Avramidi:2000bm,kirstenbook,Vassilevich:2003xt,Fursaev:2011zz}. In comparison, its application to Lorentz-violating models (or non-elliptical operators) is underdeveloped.  Some particular studies can be found in the list of references above, while the attempt of a systematic treatment of such models by the heat kernel method was initiated in \cite{Nesterov:2010yi}, followed by \cite{D'Odorico:2015yaa}. However, the derivation in \cite{D'Odorico:2015yaa} involves manipulations of asymptotic
 expansions which are hard to justify even within the analytic continuation method.
 One of the  goals of the present paper is to circumvent this difficulty by using the resolvent method for the heat kernel. In addition, we will present a number of practical tools to simplify the general but tedious procedure of \cite{Nesterov:2010yi,D'Odorico:2015yaa} for special types of anisotropic operators.

We perform all calculations in an arbitrary $(d+1)$-dimensional curved background space-time and exclusively work with covariant techniques for geometries with a preferred foliation adapted to the $(d+1)$-splitting. Our work is  organized in the following way. In section \ref{sec:summary} we introduce basic definitions, anisotropic operators and the general setup for Lifshitz models, including their extension to curved space-time. We also give a brief overview about the heat kernel method for the one-loop effective action and its ultraviolet divergences and formulate relevant anisotropic scaling properties. Section \ref{sec:general} presents the extension of the short-time expansion technique (also known in physics literature as Schwinger-DeWitt expansion \cite{Barvinsky:1985an}) to    Lorentz-violating operators of Lifshitz type in curved space-time.  We show how the operator resolvent method reduces the calculation of the Schwinger-DeWitt expansion for rather generic anisotropic Lifshitz operators to 
an algorithm involving the relativistic minimal covariant second order operator.

Section \ref{sec:calc_fo} is devoted to the development of efficient computational techniques to obtain the heat kernel coefficients of certain anisotropic operators in terms of the heat kernel coefficients of isotropic (relativistic) operators. In section \ref{sec:calcul}, these techniques are applied to derive new generic results for the case of so called ``projectable'' space-times. In particular, in section \ref{sec:examples} we derive the one-loop effective action for two families of  anisotropic operators in $2+1$ and $3+1$ dimensions. Finally, in section \ref{sec:concl}, we conclude with a summary and an outlook for future directions. Several  technical aspects are collected in appendices \ref{app:1}-\ref{sec:func}.


\section{Quantum Lifshitz theories}\label{sec:summary}

\subsection{Lifshitz theories in curved spaces}\label{sec:curved}

A Lifshitz theory on a curved space-time is a field theory which is invariant under the subgroup of diffeomorphisms that preserve a certain foliation in space-like hypersurfaces. Suppose that the space-time has dimension $(d+1)$, and let us choose coordinates $X^\mu=(x^i,t)$ (with $\mu=0,\dots d$ and $i=1,\dots,d$) such that the leaves of the foliation are the hypersurfaces of constant $t$. Then the foliation-preserving diffeomorphisms (FDiffs) take the form
\be
\label{eq:FDiff}
t\mapsto \tilde t(t), \qquad x^i \mapsto \tilde x^i(t,x^j).
\ee
The space-time metric and the coordinate system chosen uniquely define the lapse function $N(t,\mathbf{x})$, the shift vector $N^i(t,\mathbf{x})$ and the spatial metric $\gamma_{ij}(t,\mathbf{x})$ via the equation (we work with Euclidean signature)
    \be
    \label{eq:ADM}
    \di s^2\equiv g_{\mu\nu}\,\di X^\mu\,\di X^\nu=N^2c^2\,\di t^2
    +\gamma_{ij}(\di x^i+N^i\di t)(\di x^j+N^j\di t),
    \ee
where the speed of light $c$ has been included in order to make the lapse dimensionless in any system of units. We will refer to these three objects collectively as the ADM structure \cite{Arnowitt:1959ah}. Under infinitesimal FDiffs generated by a vector field
\be
f=f^0(t)\partial_t+f^i(t,\mathbf{x})\partial_i,
\ee
the ADM structure transforms as
\begin{equation}\label{eq:ADMtr}
\delta^f N={\cal L}_{{f}}N+\dot f^0N,\quad
\delta^f N^i={\cal L}_{{f}}N^i+\dot f^0N^i+\dot f^i,\quad
\delta^f\gamma_{ij}={\cal L}_{{f}}\gamma_{ij},
\end{equation}
where ${\cal L}_{ {f}}$ denotes the Lie derivative along the vector field ${f}$.
As we see, the spatial metric transforms covariantly, but the lapse and the shift do not. In particular, the second equation above tells us that the shift $N^i$ can always be brought to zero by a time-dependent spatial diffeomorphism ($f^0=0$), and the first equation implies that, if the lapse depends only on time, $N(t,\mathbf{x})=N(t)$, then it can be brought to one by a time reparameterization ($f^i=0$). We may use the ADM structure to define two derivative operators acting on spatial tensor fields, which transform covariantly under FDiffs: the spatial covariant derivative $\nabla_i$ compatible with the spatial metric,  $\nabla_i\gamma_{jk}=0$, and the covariant time derivative $D_t$, defined as
\begin{equation}
    D_t\equiv\frac1N\,\big(\,\partial_t-{\cal L}_{\vec{N}}\big),\label{eq:covDt}
 \end{equation}
 where ${\cal L}_{\vec{N}}$ denotes the Lie derivative along the shift vector $N^i$.
One can check using (\ref{eq:ADMtr}) that this operator indeed transforms covariantly under FDiffs, that is, $\delta^f(D_t\phi)={\cal L}_{{f}}(D_t\phi)$ for any spatial tensor field $\phi$ provided that $\phi$ itself transforms covariantly. Since we do not have the full diffeomorphism group at our disposal, the Riemann tensor of the space-time metric does not completely characterize its curvature, in the sense that the vanishing of that tensor does not guarantee that there will be an FDiff which maps the space-time metric to the standard Euclidean metric. The curvature is instead characterized by three spatial tensor fields: the Riemann tensor $R_{ijkl}$ of the spatial metric $\gamma_{ij}$, the extrinsic curvature $K_{ij}$ and the acceleration vector $a_i$. The latter are defined as
\be
\label{eq:Ka}
K_{ij}\equiv \frac{1}{2N}(\partial_t \gamma_{ij}-\nabla_{i}N_{j}-\nabla_{j}N_{i})=\frac12\,D_t\gamma_{ij},\qquad a_i\equiv\nabla_i\ln N.
\ee
Note from (\ref{eq:ADMtr}) and the covariance of $D_t$ and $\nabla_i$ that these three tensors transform covariantly under FDiffs. It is easy to check that, indeed, they completely characterize the curvature, namely that the vanishing of the three of them implies that there is an FDiff which maps the space-time metric to the standard Euclidean metric.

The action of a Lifshitz theory for dynamical fields $\phi^A(t,\mathbf{x})$, where $A$ is a collective index which includes spatial tensor indices and a label for different fields, is constructed with a kinetic part which is quadratic in time derivatives to ensure perturbative unitarity, while the potential part compatible with unitarity  can be rather general. A generic FDiff-invariant action functional with these properties and which is quadratic in the dynamical fields can be written in the form
\begin{equation}\label{eq:action}
S=\frac{1}{2}\int\di t\,\di^\ds x\, N\sqrt\gamma\,G_{AB}\,\phi^A { \D}^{B}_{C}\, \phi^C,
\end{equation}
where $G_{AB}(t,\mathbf{x})$ is a metric on the bundle in which the fields $\phi^A$ take values and
\begin{eqnarray}
  &&\D\equiv -\Delta_t+\frac{c^2}{M^{2(z-1)}}(-\Delta_x)^z+\sum_{m=0}^{2z-1}\Omega_{(m)}^{i_1\dots i_m}\nabla_{i_1}\dots\nabla_{i_m},\label{eq:gen_D}\\
  &&(\Delta_t)^{A}_B\equiv\frac{1}{\sqrt\gamma}\,G^{AC}D_t\left(\sqrt\gamma\, G_{CB}D_t\right),\qquad (\Delta_x)^{A}_B\equiv\frac{1}{N}G^{AC}\nabla_i\left( N G_{CB}\nabla^i\right).\label{eq:Deltatx}
\end{eqnarray}
Here, $z$ is a natural number, $M$ is a constant with dimensions of inverse length and the coefficients $[\Omega_{(m)}^{i_1\dots i_m}]^{A}_B(t,\mathbf{x})$ are generic spatial tensor fields, which represent the coupling to the background geometry, a background configuration for the dynamical fields or other types of background fields. In (\ref{eq:gen_D}) we suppressed the indices $A$ and $B$, and we will continue  suppressing them if no confusion is possible.
The operators $\Delta_t$ and $\Delta_x$ can be rewritten as
\begin{equation}
(\Delta_t)^{A}_B=G^{AC}(D_t+K)\, G_{CB}D_t,\qquad (\Delta_x)^{A}_B=G^{AC}(\nabla_i +a_i) G_{CB}\nabla^i,
\end{equation}
where $K=\gamma^{ij}K_{ij}$ is the trace of the extrinsic curvature. The structure of these operators ensures that they are Hermitian with respect to the inner product
\be\label{eq:inner}
   \langle \phi, \psi \rangle \equiv
    \int \di t\,\di^\ds x\, N\sqrt\gamma\,
    G_{AB}\, \phi^A\,\psi^B.
    \ee
We assume that the coefficients $\Omega_{(m)}$ are such that the third term in ${\D}$ is also Hermitian, so that $\D$ itself is Hermitian.
Since it is constructed out of covariant objects, the action (\ref{eq:action}) is invariant under FDiffs provided that the dynamical fields $\phi^A$ and the background fields $G_{AB}$ and $\Omega_{(m)}$ transform covariantly. In fact, although in the above discussion we have focused on infinitesimal FDiffs, the action (\ref{eq:action}) turns out to be invariant under all FDiff transformations (\ref{eq:FDiff}), not only those generated by a vector field. In particular, it is invariant under time reversal and parity. Note finally that the operator (\ref{eq:gen_D}) has the structure
\begin{equation}\label{eq:oper_form}
\D=-D_{t}^2+\frac{c^2}{M^{2(z-1)}}(-\Delta)^z+{\cal C}(\nabla,D_t),
\end{equation}
where $\Delta=\gamma^{ij}\nabla_i\nabla_j$ is the standard spatial Laplace-Beltrami operator and ${\cal C}(\nabla,D_t)$ involves no more than $2z-1$ spatial derivatives and no more than one time derivative. We will make use of this property in section \ref{sec:general}.

So far we have not chosen any particular system of units. There are two natural options: the usual units $c=1$, which we call {\emph{physical units}}, and a system of units in which the whole prefactor $c/M^{z-1}$ in (\ref{eq:gen_D}) is equal to one. We will refer to the latter as {\emph{anisotropic units}}.
Since the constant $c/M^{z-1}$ has dimensions of (length)$^z$/time, it follows that, in anisotropic units, time has dimensions of (length)$^z$. Let us find the dimensions of the different objects that appeared above in anisotropic units. First, from (\ref{eq:ADM}) we see that, in any system of units, the lapse $N$ and the spatial metric $\gamma_{ij}$ are dimensionless, whereas the shift $N^i$ has dimensions of length/time. Therefore, in anisotropic units we have
\begin{equation}
[N]=0,\quad [N^i]=z-1,\quad [\gamma_{ij}]=0,
\end{equation}
where $[a]=b$ means that $a$ has dimensions of (length)$^{-b}$.
Taking this into account it is clear that
\begin{equation}\label{eq:dim_der}
[D_t]=z,\quad [\nabla_i]=1.
\end{equation}
From the above two equations we find the anisotropic dimension of the curvatures,
\begin{equation}\label{eq:dim_curv}
[R_{ijkl}]=2,\quad [K_{ij}]=z,\quad [a_i]=1,
\end{equation}
and that of the operator (\ref{eq:gen_D}),
\begin{equation}\label{eq:dim_D}
[\D]=2z.
\end{equation}
Finally, from this equation and the second equation in (\ref{eq:dim_der}) we find the anisotropic dimension of the background fields $\Omega_{(m)}$ in (\ref{eq:gen_D}),
\begin{equation}\label{eq:dim_omega}
[\Omega_{(m)}]=2z-m.
\end{equation}
Note that both systems of units, physical and anisotropic, coincide in the case $z=1$. Therefore, the physical dimension of all these objects is obtained by setting $z=1$ in the above equations. We will use anisotropic units throughout this paper except in section \ref{sec:general}, where we will switch to physical units.

Our general method, to be described below, may be applied and extended to anisotropic operators different from \eqref{eq:gen_D}, but for concreteness we focus on the family of operators \eqref{eq:gen_D} which already encompasses many interesting and physically relevant situations.
 Throughout this paper we work in manifolds without boundaries or with trivial boundary conditions--- asymptotically flat with zero boundary conditions at infinity.
%

\subsection{Effective action from the heat kernel}\label{sec:eff_heat}
We compute the quantum one-loop effective action for the action \eqref{eq:action} by using heat kernel (Schwinger-DeWitt) techniques \cite{DeWittbook:1965,Gilkeybook:1984,Barvinsky:1985an} (see  \cite{Avramidi:2000bm,kirstenbook,Vassilevich:2003xt,Fursaev:2011zz,
Barvinsky-Scholarpedia} for  reviews in the relativistic case). The one-loop effective action is given by {\footnote{A naive evaluation of the Gaussian path integral that defines the effective action for (\ref{eq:action}) gives
\begin{equation*}
W=\frac12\,{\rm Tr}\,\ln\,(G_{AC}{\cal D}^{C}_B)=\frac12\,{\rm Tr}\,\ln\,{\cal D}^A_B+\frac12\,{\rm Tr}\,\ln\,G_{AB}.
\end{equation*}
However, this misses a factor $(\det G_{AB})^{1/2}$ which is present in the path integral measure. This factor cancels the second term above, thereby yielding (\ref{eq:effec_act_3}).
}}
\be
\label{eq:effec_act_3}
W=\frac{1}{2}\,\Tr \ln {\cal D},
\ee
where $\D$ is the operator defined in (\ref{eq:gen_D}).
This quantity is divergent and requires renormalization. We perform the renormalization via the zeta-function method \cite{Dowker:1975tf,Hawking:1976ja}
\be
\label{eq:regac}
W=-\frac{1}{2}\left.\frac{\di}{\di s} \Tr\,\,\frac{\mu^{2s}}{{\cal D}^s}\,\right|_{\;s=0}=-\frac{1}{2}\zeta_{\cal D}'(0)-\frac{1}{2}\log \m^2 \zeta_{\cal D}(0),
\ee
where we have introduced a scale $\mu$ with the physical dimension of energy, which reflects the renormalization ambiguity. The generalized zeta function $\zeta_\D(s)$ is defined as
\be
\label{eq:zeta}
 \zeta_{\cal D}(s)\equiv \Tr \,{\cal D}^{-s}= \frac1{\Gamma(s)}\int_0^\infty \di \tau\, \tau^{s-1}\,\Tr (e^{-\tau \D})
\ee
for those values of $s$ at which this expression converges, and by analytic continuation elsewhere.
In particular, $\zeta_\D(s)$ is well-defined near $s=0$, so that the renormalized effective action (\ref{eq:regac}) is finite.

Eqs.~(\ref{eq:regac}) and (\ref{eq:zeta}) turn the problem of computing the effective action into the problem of computing the trace of the operator $e^{-\tau\D}$. This operator is called the heat operator, and its kernel $H_\tau$, defined by
\begin{equation}\label{eq:heatop-k}
(e^{-\tau\D}\phi)^{A}(t,\mathbf{x})=\int\di t'\, \di^\ds x'\,N\sqrt\gamma\,(H_\tau)^{A}_B(t,\mathbf{x};t',\mathbf{x}')\phi^B(t',\mathbf{x}'),
\end{equation}
is called the heat kernel. The heat kernel is the unique solution to the initial value problem
\begin{equation}\label{eq:heateq}
(\partial_\tau+\D)H_\tau=0,\quad (H_0)^{A}_B(t,\mathbf{x};t',\mathbf{x}')=\delta^{A}_B\,\delta^{(d+1)}(t,\mathbf{x};t',\mathbf{x}'),
\end{equation}
where the delta function appearing in the initial condition is dedensitized in both arguments. The differential equation above is analogous to the heat equation, hence the terminology. In terms of the heat kernel we have for any function\footnote{The generic divergences associated to volume effects can be regularized by assuming proper fall-off conditions on  the smearing function $F$ \cite{Vassilevich:2003xt}. 
They have no impact on the short-distance properties of the  formalism.} $F(t,\mathbf{x})$
\begin{equation}\label{eq:trace_h}
\Tr (Fe^{-\tau \D})=\int\di t\,\di^\ds x\, N\sqrt\gamma\,F\, (H_\tau)^{A}_A(t,\mathbf{x};t,\mathbf{x}).
\end{equation}
Note that the integrand in \eqref{eq:trace_h} is a scalar function on space-time which is constructed out of the same background fields as $\D$, and which has the same symmetries as $\D$.

Let us see what else we can say about the trace (\ref{eq:trace_h}), before doing a explicit calculation. We will assume for simplicity that $\D$ acts on multiplets of scalar fields and that the bundle metric $G_{AB}$ is just a constant matrix, so that it disappears from $\D$ (see (\ref{eq:gen_D})). Our main conclusions, however, remain unaltered in the general case. The initial value problem (\ref{eq:heateq}) is easily solved (using Fourier methods) in the case where the space-time metric is flat and the coefficient fields $\Omega_{(m)}$ vanish. For general backgrounds, it can be solved iteratively taking this ``flat'' solution as the zero-th iteration. The result of this iterative procedure is that the function $(H_\tau)^{A}_A$ can be expressed as an expansion in powers of the curvatures $R_{ijkl}$, $K_{ij}$ and $a_i$, the fields $\Omega_{(m)}$ and their derivatives. A generic term in this expansion is a scalar of the schematic form
\begin{equation}\label{eq:gen_term}
D_{t}^u\nabla^vR^pK^qa^r\prod_{m=0}^{2z-1}\Omega_{(m)}^{s_m},
\end{equation}
i.e. containing $u$ time derivatives, $v$ spatial derivatives, $p$ factors of the spatial Riemann tensor, $q$ powers of the extrinsic curvature, $r$ powers of the acceleration and
$s_m$ powers of the fields $\Omega_{(m)}$. The previous terms are multiplied by a coefficient which depends on $d$, $z$ and  $\tau$. Note that the natural numbers $u$, $v$, etc.~in (\ref{eq:gen_term}) are not completely arbitrary. Indeed, by time-reversal invariance the total number of time derivatives must be even. This number is $u+q$, because there is a time derivative hidden in the extrinsic curvature $K_{ij}$, so we have
\begin{equation}\label{eq:Tconstraint}
u+q\in 2{\mathbb N}.
\end{equation}
On the other hand, the total number of indices in (\ref{eq:gen_term}) must be even so that they can be contracted to form a scalar. The number of indices of $D_t$, $R_{ijkl}$ and $K_{ij}$ is already even, so this condition implies
\begin{equation}\label{eq:sconstraint}
v+r+\sum_{m=0}^{2z-1}m\,s_m\in 2{\mathbb N}.
\end{equation}
We may determine how the coefficient in front of (\ref{eq:gen_term}) depends on $\tau$ by dimensional analysis. Indeed, let us choose anisotropic units. Since the heat operator $e^{-\tau\D}$ is necessarily dimensionless, it follows from (\ref{eq:heatop-k}) that the heat kernel $H_\tau$ has anisotropic dimension $d+z$. On the other hand, by (\ref{eq:dim_der}), (\ref{eq:dim_curv}) and (\ref{eq:dim_omega}) the anisotropic dimension of the scalar (\ref{eq:gen_term}) is the natural number
\begin{equation}\label{eq:dim_term}
n=z(u+q)+2p+v+r+\sum_{m=0}^{2z-1}(2z-m)s_m.
\end{equation}
In order for the dimensions to match, the coefficient in front of (\ref{eq:gen_term}) (let us call it $\alpha$) must satisfy $[\alpha]=d+z-n$. Now, the operator $\D$ in (\ref{eq:gen_D}) has no dimensionful parameters in anisotropic units, so the only dimensionful quantity we have at our disposal is $\tau$. Since the product $\tau\D$ must be dimensionless, Eq.~(\ref{eq:dim_D}) implies $[\tau]=-2z$, and therefore $\alpha\propto \tau^{(n-d-z)/2z}$. In other words, what we have found is that the coincidence limit of the heat kernel is given by the asymptotic expansion in $\tau\to 0$,
\begin{equation}\label{eq:hexpansion}
(H_\tau)^{A}_A(t,\mathbf{x};t,\mathbf{x})=\frac{1}{\tau^{(d+z)/2z}}\sum_{n=0}^\infty b_n[\D]\tau^{n/2z},
\end{equation}
where $b_n[\D]$ is a linear combination of scalars of the form (\ref{eq:gen_term}) which satisfy (\ref{eq:dim_term}), with coefficients that are independent of $\tau$, the background fields and the space-time point (and hence may only depend on $d$ and $z$). In fact, (\ref{eq:Tconstraint}) and (\ref{eq:sconstraint}) imply that the right-hand side of (\ref{eq:dim_term}) is even, so that
\begin{equation}\label{eq:oddn}
b_n[\D]=0\quad{\text{for }}n\in 2{\mathbb N}+1.
\end{equation}
We stress that in the case of manifolds with boundary and non-trivial boundary conditions there are extra contributions to the heat kernel coefficients. 
Substituting (\ref{eq:hexpansion}) into (\ref{eq:trace_h}) we obtain
\begin{equation}
    \label{eq:asymp}
    \Tr ( F e^{-\tau \D})    = \frac{1}{\tau^{(d+z)/2z}}
    \sum_{n=0}^\infty B_n[F,\D]\, \tau^{n/2z},
    \end{equation}
where
\begin{equation}\label{eq:bintegrands}
B_n[F,\D]=\int\di t\,\di^\ds x\, N\sqrt\gamma\,F\,b_n[\D].
\end{equation}
What remains to be done in order to compute $\Tr ( F e^{-\tau \D})$ is to find the explicit value of the numerical coefficients appearing in $b_n[\D]$, and hence also in $B_n[F,\D]$. A general method for computing these coefficients, together with a more explicit derivation of (\ref{eq:asymp}), will be presented in section \ref{sec:general} (see also \cite{Nesterov:2010yi,D'Odorico:2015yaa,Baggio:2011ha}). Note that the above argument is valid in particular for $z=1$. In that case, the operator (\ref{eq:gen_D}) can be tuned so that the action (\ref{eq:action}) is relativistic, namely invariant under the full diffeomorphism group. Different techniques for computing the heat kernel coefficients in the relativistic case can be found e.g.~in \cite{fegan1985,Fursaev:2011zz,
Birrell:1982ix,Gilkey:2004dm,Avramidi:2000bm,Barvinsky:1985an,DeWitt:2003pm}.

Let us return to the one-loop effective action. Given the asymptotic expansion \eqref{eq:asymp}, the analytic continuation from (\ref{eq:zeta}) gives \cite{kirstenbook}
\be
\label{eq:zeta0}
\zeta_{\cal D}(0)=B_{\ds+z}[\,1,\D\,],
\ee
so that by \eqref{eq:regac} we have
\begin{align}
W=-\frac{1}{2}\zeta'_{\cal D}(0)-\frac{1}{2}\log \m^2 B_{\ds+z}[\,1,\D\,].\label{eq:effact}
\end{align}
We will mainly be interested in the computation of the coefficient $b_{\ds+z}[\D]$, which from (\ref{eq:bintegrands}) and (\ref{eq:effact}) gives the dependence of the effective action on the renormalization scale $\mu$. From (\ref{eq:oddn}) we see that this coefficient vanishes if $d+z$ is odd, so in this case the effective action has no $\mu$ dependence.
We may thus conclude that {\emph{Lifshitz theories in $\ds+1$ dimensions with a degree of anisotropy $z$, such that $\ds+z$ is odd, are one-loop finite}}. This is the anisotropic generalization of the well-known result that relativistic theories (corresponding to $z=1$) are one-loop finite in odd space-time dimension. Note also that we can gain information about the structure of $b_{d+z}[\D]$ by setting $n=d+z$ in (\ref{eq:dim_term}) and using (\ref{eq:Tconstraint}). For example, in the case $z>d$ these equations imply that $b_{d+z}[\D]$ involves no time derivatives of the background fields, and in the case $z=d$ they imply that the only possible terms involving time derivatives are $K_{ij}K^{ij}$, $K^2$ and $D_tK$, times a constant that may only depend on $d$. We will make crucial use of this information in section \ref{sec:calcul}.

\section{General technique for the heat kernel operator}\label{sec:general}

The calculation of the coefficients $B_n[\,F,{\cal D}\,]$ in the asymptotic expansion (\ref{eq:asymp})
for the most general case can be performed with the aid of the Schwinger-DeWitt technique \cite{Barvinsky:1985an}. The latter  is directly applicable only to a special class of second-order differential operators -- {\em minimal} operators whose derivatives form the d'Alembertian (space-time Laplacian). Extension of this technique to generic higher derivative operators and operators of Lifshitz theories requires special methods which we describe here.

\subsection{Minimal second order operators vs Lifshitz operators}\label{sec:minimal}

We remind that our Lifshitz type operator ${\cal D}(\nabla)$ acts in the $(d+1)$-dimensional space-time with coordinates $X^\mu=(x^i,t)$, $i=1,2,...d$. Now let us consider the operator of second order in Riemannian covariant derivatives $\nabla_\mu$ relative to the space-time metric $g_{\mu\nu}$ of {\em minimal} form
    \begin{eqnarray}
    {\cal D}(\nabla_X)=-g^{\mu\nu}\nabla_\mu\nabla_\nu+P(X). \label{minimal}
    \end{eqnarray}
For such an operator one can write down an asymptotic expansion of its heat kernel at separated points $X\neq Y$ at $\tau \to 0$ with coefficients $a_n(X,Y)$ satisfying a well-known set of recurrent equations   \cite{DeWittbook:1965,Barvinsky:1985an}. This expansion has the form\footnote{The Van Vleck determinant here is dedensitized and the kernel as well as the delta function in (\ref{kernel}) are densities of zero weight in $X$  and unit weight in the second argument $Y$, which explains the origin of the factor $g^{1/2}(Y)$ in (\ref{kernel}).}
    \begin{eqnarray}
    &&e^{-\tau{\cal D}(\nabla_X)}\delta^{(d+1)}(X,Y)=
    \frac{\Delta^{1/2}(X,Y)}{(4\pi\tau)^{(d+1)/2}}\,
    \, g^{1/2}(Y)\,
    e^{-\sigma(X,Y)/2\tau}\,
    \sum\limits_{n=0}^\infty a_n(X,Y)\,\tau^n, \label{kernel}
    \end{eqnarray}
where $g(X)={\rm det}\,g_{\mu\nu}(X)$, $\sigma(X,Y)$ and $\Delta^{1/2}(X,Y)$  are the Synge world function and its Pauli-Van Vleck determinant \cite{DeWittbook:1965},
    \begin{eqnarray}
    \Delta(X,Y)=g^{-1/2}(X)\,g^{-1/2}(Y)\,{\rm det}\,\frac{\partial^2\sigma(X,Y)}{\partial X^\mu\partial Y^\nu}.
    \end{eqnarray}
The quantities $a_n(X,Y)$ are the two-point Schwinger-DeWitt coefficients built in terms of the metric $g_{\mu\nu}$, its space-time curvature, the potential term $P$ of the operator and their covariant derivatives (\ref{minimal}). When the operator ${\cal D}(\nabla)={\cal D}^A_B(\nabla)$ acts in the space of columns of fields $\phi^A(X)$ labeled by generic  indices $A$, the coefficients $a_{n}(X,Y)=a_{n\,B}^A(X,Y)$ represent the corresponding matrices in field space. For separate points in $a_n(X,Y)$ these coefficients are nonlocal, but the recurrent equations for them allow one to prove that their coincident limits $a_n(X,X)$ are local and explicitly calculable.

The  general statement about  the large class of differential operators considered in the last section, Eq.~(\ref{eq:asymp}), matches with the expansion (\ref{kernel}) for the case $z=1$ corresponding to the minimal second order operator (\ref{minimal}). In this case,
the quantities $B_n[\,F,{\cal D}\,]$ are related to the Schwinger-DeWitt coefficients $a_n$ in (\ref{kernel}) by,
    \begin{eqnarray}
    B_{2n}[\,F,\D(\nabla_X)]=\frac1{(4\pi)^{(d+1)/2}}\int \di^{d+1}X\, g^{1/2}(X)\,{\rm tr} \big[\,F(X)\,a_n(X,X)\,\big],\quad z=1.
    \end{eqnarray}
Odd  coefficients $B_{2k+1}[\,F,{\cal D}(\nabla_X)]=0$  vanish in space-times without a boundary (or asymptotically flat spaces with zero boundary conditions at infinity). In case of boundaries they are given by local boundary surface integrals which are not explicitly related to the Schwinger-DeWitt coefficients but can be extracted from them  \cite{McKean-Singer, Barvinsky-Nesterov}.

Our general strategy for the calculation of $B_{\ds+z}[\,F,\D\,]$ for a Lifshitz type operator will be to express it in terms  of the properties of minimal operator of the above type. The Lifshitz  operator of the form \eqref{eq:oper_form} (see also \eqref{eq:gen_D}) implies the following decomposition
    \begin{eqnarray}
    &&{\cal D}={\cal A}+G({\cal B})+{\cal C},\quad
    {\cal A}=-D_t^2,\quad {\cal B}=-\Delta       \label{AB}
    \end{eqnarray}
with the function
    \begin{eqnarray}
    G(q)=\frac{q^z}{M^{2(z-1)}}.   \label{G(q)}
    \end{eqnarray}
This decomposition isolates the leading symbol of the operator with respect to time derivatives in ${\cal A}$ and the leading symbol with respect to spatial derivatives --- in $G({\cal B})$. The operator ${\cal C}(D_t,\nabla)$, which is of the order $2z-1$ in space derivatives and linear in first order time derivatives, is subleading to ${\cal A}$ and $G(-\Delta)$. Though we will work with this particular choice of ${\cal A}$, ${\cal B}$, other choice of these operators maintaining this order of derivatives is also possible. In particular, both ${\cal A}$ and ${\cal B}$ might be chosen as to include terms with first order space-time derivatives. The results can also be generalized to a function $G(q)$ polynomial in $q$ and starting with the leading term (\ref{G(q)}). For these reasons we begin with a generic situation and  only afterwards restrict ourselves to the case of (\ref{AB}) and (\ref{G(q)}).

\subsection{The resolvent method for general operators} \label{sec:AGB}

The idea of the method is to reduce the calculation for the principle part of ${\cal A}+G({\cal B})$ to the case of a linear combination of ${\cal A}$ and ${\cal B}$, while treating the lower derivative ${\cal C}$ parts as perturbations. Therefore, the efficient treatment is possible when the asymptotic expansion for the heat-kernel of the linear combination of ${\cal A}$ and ${\cal B}$ is known. We follow the approach suggested in \cite{Nesterov:2010yi} and complete it by providing a rigorous treatment of all the steps which was so far missing (see also \cite{D'Odorico:2015yaa} for a related proposal). For this we introduce the  operator ${\cal M}(\cal A,B)$ which allows one to break the exponentiated sum of two arbitrary non-commuting operators ${\cal A}$ and ${\cal B}$ into the product 
    \begin{eqnarray}
    &&e^{\cal A+B}=e^{\cal A}\,e^{\cal B}\,{\cal M}({\cal A,B}),
    \quad {\cal M}({\cal A,B})=1+
    O\big([\,{\cal A,B}\,]\big),
    \end{eqnarray}
the Zassenhaus formula, whose coefficients can efficiently be calculated to a high order  \cite{CPA:CPA3160070404}. The inverse of this operator ${\cal M}^{-1}({\cal A,B})$  can be used to write
    \begin{eqnarray}
    \label{eq:exp_tog}
    &&e^{\cal A}\,e^{\cal B}=e^{\cal A+B}\,{\cal M}^{-1}({\cal A,B})\,.
    \end{eqnarray}
\begin{figure}
  \centering
  \includegraphics[scale = 0.3]{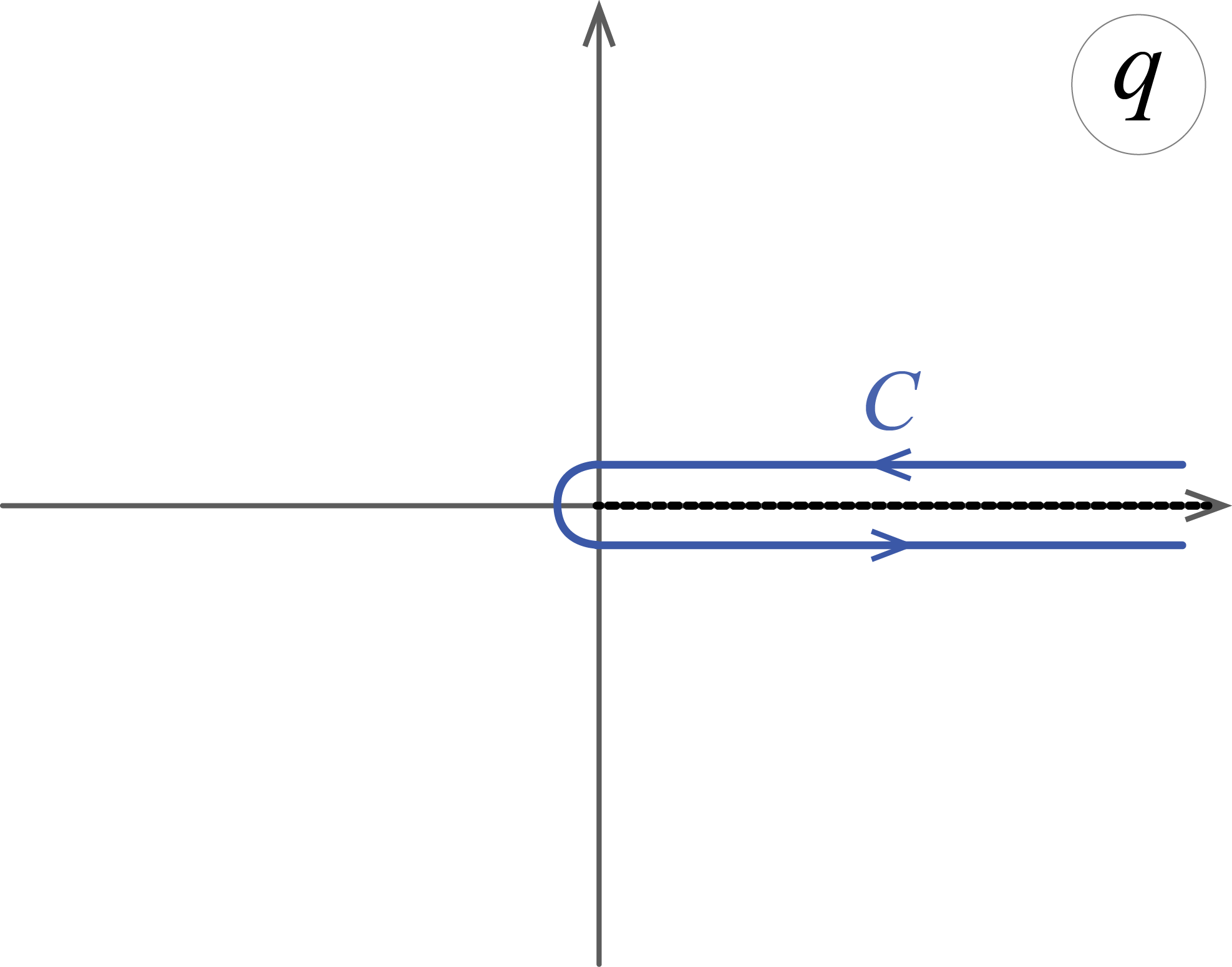}
  \caption{Contour $C$ in $q$ plane.}\label{fig_contour_C}
\end{figure}
With this definition we have
    \begin{eqnarray}
    e^{-\tau{\cal A}-\tau G({\cal B})-\tau{\cal C}}
    =e^{-\tau{\cal A}}\,e^{-\tau G({\cal B})}\,{\cal M}(-\tau{\cal A},-\tau G({\cal B}))\,e^{-\tau{\cal C}}
    {\cal M}(-\tau{\cal A}-\tau G({\cal B}),-\tau{\cal C}).          \label{expFplusBox}
    \end{eqnarray}
We consider the  representation for the factor $e^{-\tau G({\cal B})}$ in terms of the contour integral of the resolvent of $\cal B$,
    \begin{eqnarray}
    e^{-\tau G({\cal B})}=\frac1{2\pi i}\int\limits_C
    \di q\,\frac{e^{-\tau G(q)}}{q-{\cal B}}=
    -\frac1{2\pi i}\int\limits_C \di q\!\!
    \int\limits_0^{i\,{\rm sign}({\rm Im}\,q)\infty}\!\! \di \lambda\;
    e^{-\tau G(q)+\lambda(q-{\cal B})},    \label{Fourier}
    \end{eqnarray}
with  the contour $C$ encircling the spectrum of $\cal B$. We only work with ${\cal B}$ being positive semidefinite and thus this contour encircles the semi-axis of real positive values of $q$ with the contour running counterclockwise  (Figure \ref{fig_contour_C}). The resolvent $1/(q-\cal B)$ of the operator $\cal B$,  is represented in (\ref{Fourier}) by the integral over the auxiliary parameter $\lambda$ with the upper limit $i\,{\rm sign}\,[{\rm Im}\,(q-{\cal B})]\infty=i\,{\rm sign}({\rm Im}\,q)\infty$, which provides the convergence of the integral at imaginary infinity of the complex $\lambda$-plane.

Substituting \eqref{Fourier} back into (\ref{expFplusBox}), we use \eqref{eq:exp_tog} to recombine the two operator factors into one heat kernel $e^{-\tau{\cal A}-\lambda{\cal B}}$. Using the cyclicity of the trace, the final result for the heat kernel trace is
    \begin{eqnarray}
    &&{\rm Tr}\,e^{-\tau{\cal D}}=
    -\frac1{2\pi i}\int\limits_C \di q\!\!
    \int\limits_0^{i\,{\rm sign}({\rm Im}\,q)\infty}\!\! \di \lambda\;\,
    e^{-\tau G(q)+\lambda q}\;\,{\rm Tr}\,\Big(
    \varOmega\,e^{-\tau{\cal A}-\lambda{\cal B}}\Big),     \label{1000}
    \end{eqnarray}
with
      \be
   \varOmega=
    {\cal M}^{-1}(-\tau{\cal A},-\lambda{\cal B})\,
    {\cal M}(-\tau{\cal A},-\tau G({\cal B}))\,e^{-\tau{\cal C}}
    {\cal M}(-\tau{\cal A}-\tau G({\cal B}),-\tau{\cal C}).   \label{total_Zassenhaus}
   \ee

Let us now specialize to the operators ${\cal A}=-D_t^2$ and ${\cal B}=-\Delta$, defined in \eqref{AB}. Their combination with numerical coefficients $\lambda$ and $\tau$ forms a minimal second order operator $\tilde{\cal A}$
    \begin{eqnarray}
    &&\lambda{\cal B}+\tau{\cal A}=
    \lambda\left(\frac\tau\lambda\,n^\mu n^\nu+\gamma^{ij}\delta^\mu_i\delta^\nu_j\right)
    \partial_\mu\partial_\nu+... \equiv\lambda\, \tilde {\cal A}   \,, \label{tilde_g}
    \end{eqnarray}
in which second order derivatives are contracted with the {\em generalized} contravariant metric $\tilde g^{\mu\nu}=(\tau/\lambda)\,n^\mu n^\nu+\gamma^{ij}\delta^\mu_i\delta^\nu_j$ and the ellipsis denote lower derivative terms. The corresponding covariant metric $\tilde g_{\mu\nu}$ reads
    \be
    \di\tilde s^2=\tilde N^2\di t^2
    +\gamma_{ij}\,(\di x^i+N^i \di t)\,(\di x^j+N^j \di t),\quad
    \tilde N=\sqrt{\frac{\lambda}{\tau }}\,N.  \label{eq:ADM_t}
    \ee
and it can be used to construct covariant space-time derivatives $\tilde\nabla_\mu$ compatible with this metric, so that the linear combination (\ref{tilde_g}) takes the form of the minimal differential operator covariant with respect to $\tilde g_{\mu\nu}$, which parametrically depends on the ratio $\tau/\lambda$,
    \begin{eqnarray}
    &&\lambda{\cal B}+\tau{\cal A}=\lambda\,
    {\cal A}(\tilde\nabla,\tau/\lambda),\quad
    {\cal A}(\tilde\nabla,\tau/\lambda)=\tilde g^{\mu\nu}\tilde\nabla_\mu\tilde\nabla_\nu+\tilde P.
    \end{eqnarray}

The asymptotic expansion of this operator for $\lambda\to 0$ at fixed $\tau/\lambda$ has the standard form (\ref{kernel}),
    \begin{eqnarray}
    &&e^{-\lambda\tilde{\cal A}(\nabla_X)}
    \delta^{(d+1)}(X,Y)=
    \frac{\tilde\Delta^{1/2}(X,Y)}{(4\pi\lambda)^{(\ds+1)/2}}\,
    \tilde g^{1/2}(Y)\,
    e^{-\tilde\sigma(X,Y)/2\lambda}\,
    \sum\limits_{n=0}^\infty \tilde a_n(X,Y)\lambda^n, \label{tildekernel}
    \end{eqnarray}
with all the objects defined with respect to the new metric $\tilde g_{\mu\nu}$ and the potential term $\tilde P(X)$.

From its definition \eqref{total_Zassenhaus}, it is clear that the operator   $ \varOmega$, when it is expanded in powers of all three operators ${\cal A}$, $\cal B$, $\cal C$ and their commutators,  is an infinite series of the following form
   \begin{eqnarray}
   &&\varOmega=\sum\limits_{k,r,s,l,p\geq 0}
   \mathfrak{B}_{(k)}\frac{\tau^r\,
   \lambda^s}{M^{2p(z-1)}}D^l,                      \label{Omega}\\
   &&2r+2s+2p(z-1)=l+k,                             \label{range}
   \end{eqnarray}
where $\mathfrak{B}_{(k)}$ is the background field quantity of the physical dimensionality $k$ built of the curvatures, their derivatives and other background quantities, $D^l$ denotes space-time derivatives $D\equiv (D_t,\nabla_i)$ of the total power $l$ and the powers of the mass parameter in the denominator follow from the structure of $G({\cal B})$ and $\cal C$ in (\ref{AB}). The restriction on the range of summation parameters follows from the dimensionless nature of the total operator $\varOmega$.

The action of (\ref{Omega}) on the heat kernel (\ref{tildekernel}) under the functional trace sign leads to the sequence of local terms originating from the coincidence limits of multiple derivatives in time or space of the Synge world function and Schwinger-DeWitt coefficients. When they are rewritten as functions of the original metric the functional trace takes the form of the series of various powers of $\lambda$ and $\tau$
   \begin{eqnarray}
   &&{\rm Tr}\,\Big(\varOmega\,e^{-\tau{\cal A}-\lambda{\cal B}}\Big)=
   \frac1{(4\pi\lambda)^{(\ds+1)/2}}\left(\frac\lambda\tau\right)^{1/2}
   \sum\limits_{n=-\infty}^\infty\,
   \sum\limits_{\;\;m\geq 0}\,
    A_{n,m}\,\lambda^n\,\tau^m.   \label{structure}
   \end{eqnarray}
Let us pause to explain the structure of this expression.
The coefficients $A_{n,m}$  are  integrals of local invariants  under FDiff \eqref{eq:FDiff}  of physical dimensionality $2(n+m)$. They are built in terms of the ADM $(d+1)$-decomposition of the original metric and other background quantities if they exist. Thus, they contain powers of the spatial $d$-dimensional curvature, acceleration of the foliation and extrinsic curvatures (collectively denoted by $R, a$ and $K$), spatial covariant derivatives $\nabla$, time reparametrization invariant derivatives $D_t$, background quantities $\Omega_{(m)}$ and $M$. Schematically,
   \begin{eqnarray}
   A_{n,m}=\int \di t\,\di^d x\,\sqrt{\gamma}\, N\!
   \sum\limits_{k,l,p,r,s,q,s_m}\frac1{M^{2k}}\!\!
   \stackrel{r+p+2l+q+s-2k+\sum_m s_m(2-m)=2(n+m)}
   {\overbrace{D_t^r\,\nabla^p\,
   R^l\, a^{q} K^s \prod_{m}^{2z-1}\Omega_{(m)}^{s_m} }}.      \label{Anm}
   \end{eqnarray}
The overall factor $(\lambda/\tau)^{1/2}$ in \eqref{structure} comes from the rescaling of $\tilde g^{1/2}$ in \eqref{tildekernel}, which generates the volume element in \eqref{Anm}. All the powers in the sum  for $\lambda$ and $\tau$ are integer, but the powers of $\lambda$ can be negative and extend to $-\infty$. This is a consequence of several facts: first, the only other appearance of $\tilde N$ is inside time derivatives and $\tilde K_{ij}$, both of which have factors of $(\tau/\lambda)^{1/2}$, cf. \eqref{eq:K_sca}. Invariance under $t\mapsto -t$ makes this factor appear always squared. Furthermore, the result of the action of the pseudo-differential operators $\varOmega$ eventually generates an infinite series in derivatives (\ref{Omega}). When acting upon the exponential factor of (\ref{tildekernel}) this generates inverse powers of $\lambda$. For instance, if the derivative $D_X\equiv (D_t,\nabla_i)$ acts $l$ times, this gives the highest possible power of $1/\lambda$, $[\,l/2\,]$ -- the integer 
part of $l/2$,
   \begin{eqnarray}
   D^l_X\,
   e^{-\sigma(X,Y)/2\lambda}\,
   \Big|_{\,Y=X}\sim\lambda^{-[l/2]},              \label{nabla_exp}
   \end{eqnarray}
and therefore generates a set of powers of $\lambda$ extending to minus infinity\footnote{Recall that only objects with two or more derivatives acting on $\sigma(X,Y)$ are non-vanishing.}. The resulting local invariants $A_{n,m}$ can have  negative physical dimensionality, which is possible thanks to the presence of the dimensionful parameter $M$ in (\ref{AB}) and the dimensions of $\Omega_{(m)}$.

 The expression in \eqref{structure} represents a double asymptotic expansion in $\tau\to 0$ and $\lambda\to 0$ which can be integrated according to  \eqref{1000}
 term by term with respect to $\lambda$ and $q$ in order to get a resulting asymptotic expansion in the remaining parameter $\tau\to 0$ \footnote{This is analogous to  the proper time integration of the heat kernel in theories with a mass parameter, when the term by term integration over $\tau$ yields the asymptotic expansion in inverse powers of a large mass-squared parameter [48]. Since we care about the UV properties of the integrals (large $q$) the role of the large mass as regulator of integrals in $\tau$ is now played by $q$.}
    \begin{eqnarray}
   {\rm Tr}\,e^{-\tau{\cal D}}=-\frac1{(2\pi i)(4\pi)^{\frac{\ds+1}{2}}}
      \sum\limits_{n=-\infty}^\infty
   \sum\limits_{m\geq 0}\, A_{n,m}\tau^{m-1/2}
  \int\limits_C \di q\,e^{-\tau G(q)}
    \int\limits_0^{\pm i\infty}\!\di \lambda\,
    e^{\lambda q}\lambda^{n-\ds/2}.               \label{expansion1}
    \end{eqnarray}

This expression, however, presents a number of difficulties. First, the  integration over $\lambda$ at zero cannot be regulated by any sort of dimensional regularization, since no finite value of space-time dimension ($d+1$)
 can render integration in all terms of \eqref{expansion1} convergent -- the sort of the problem one encounters in the Mellin transform method of \cite{D'Odorico:2015yaa}. 
 In the resolvent method we are following, this problem is circumvented by the arguments presented in App.~\ref{app:int} and gives the following simple answer which is valid for all $n$ in the range of summation\footnote{This formula and the previous manipulations can be verified by relating them to the heat kernel of
 the operator $G(\Delta)=(-\Delta)^z/M^{2(z-1)}$ and using the Fegan-Gilkey lemma  \eqref{eq:Omn},  to be discussed momentarily, to find the latter.}
    \begin{eqnarray}
    \frac1{2\pi i}\int\limits_C \di q\,e^{-\tau G(q)}
    \int\limits_0^{\pm i\infty}\!\di \lambda\;\,
    e^{\lambda q}\;\lambda^{n-d/2}=
    -\frac1{(4\pi)^{\ds/2}}\,
    \left(\frac{(M^2\tau)^{1/z}}{M^2}\right)^{n-d/2}
    \frac{\Gamma\big(\frac{d-2n}{2z}\big)}
    {z\,\Gamma\big(\frac{d-2n}{2}\big)},    \label{lambda_q_int}
    \end{eqnarray}
so that (\ref{expansion1}) takes the form
  \begin{eqnarray}
  {\rm Tr}\,e^{-\tau{\cal D}}=\frac1{(4\pi\tau)^{(\ds+1)/2}}
  \sum\limits_{n=-\infty}^\infty\,
   \sum\limits_{\;\;m\geq 0}\, A_{n,m}\tau^{n+m}
    \,\big(M^2\tau\big)^{(d/2-n)(1-1/z)}
    \frac{\Gamma\big(\frac{d-2n}{2z}\big)}
    {z\,\Gamma\big(\frac{d-2n}{2}\big)}.          \label{expansion_final}
    \end{eqnarray}

The second difficulty is that the curvature expansion of the heat kernel can be inefficient for obtaining its asymptotic expansion at small $\tau$. Indeed, in view of simple dimensional arguments a given power of $\tau$ in the expansion of the heat kernel can include an infinite series of terms of the form $(R/M^2)^n$ for arbitrary high powers of $n$. Moreover, it looks that this expansion is not analytic in $\tau\to 0$  because it has {\it a priori} an infinite number of negative powers of $\tau$. This is however not the case: as we show in App.~\ref{app:andrei}, the negative powers of $\lambda$ always appear with sufficient positive powers of $\tau$ to get rid of the negative summands  in \eqref{expansion_final}. Finally, one may be worried that the presence of $M$ and $\Omega_{(m)}$ would allow for an infinite number of operators to contribute at each order in $\tau$. The same proof that shows that there are only positive powers of $\tau$ also implies that there is a finite number of operators for a 
certain power.

Thus, the final result, which we derive in \eqref{expansion_final_2}, is that the previous expansion takes the form (\ref{eq:asymp})
   \begin{eqnarray}
    &&{\rm Tr}\,e^{-\tau{\cal D}}
    =\frac1{\tau^{\frac{\ds+z}{2z}}}
    \sum\limits_{n=0}^\infty
    \tau^{\frac{n}{z}}\,B_{2n}[1,{\cal D}],               \label{expansion_final_3}\\
    &&B_{2n}[1,{\cal D}]=\frac1{(4\pi)^{\frac{\ds+1}{2}}}\,
    \sum\limits_{m=0}^{2n} M^{\frac{2n+dz-d}z-m}\,
   B_{m,2n}[1,{\cal D}],
    \end{eqnarray}
where the integrand of $ B_{m,2n}[1,{\cal D}]$ is a {\em finite} sum of geometrical invariants of the physical dimensionality $m$.
 Summation runs over $B_k[1,{\cal D}]$ with even $k$, since coefficients with odd $k$ vanish in space-times without boundaries.

The critical coefficient $B_{d+z}$ in this expansion is given for even $d+z$ by the following sum,
   \begin{eqnarray}
    &&B_{d+z}[1,{\cal D}]=\frac1{(4\pi)^{\frac{\ds+1}{2}}}\,
    \sum\limits_{m=0}^{d+z} M^{d+1-m}\,B_{m,d+z}[1,{\cal D}], \label{critBfinal}
    \end{eqnarray}
(for odd $d+z$ it is just zero for reasons mentioned above). Since the anisotropic scaling dimension of the mass parameter $M$ is zero, in Lifshitz models most of the terms built purely of spatial curvature invariants (without time derivatives or $\Omega_{(m)}$) are vanishing, because for such invariants the physical dimensionality $m$ coincides with the scaling dimensionality. Therefore,  purely spatial curvature terms here are exhausted by the lowest order in $M$ contribution, $B_{d+z,d+z}/M^{z-1}$. Similar conclusions can be drawn for the contributions combining $\Omega_{(m)}$ and spatial curvatures.

\section{Simplification methods for special heat kernels}\label{sec:calc_fo}

In this section, we collect several methods that simplify the calculation of heat kernel traces. All these cases are covered by the general method developed in section~\ref{sec:general}, but they allow for efficient calculations for particular operators, as compared to the previous tedious algorithm.  In the case of covariant
operators, it is standard to use {\it functorial}  properties to simplify the calculation of heat kernel coefficients \cite{Vassilevich:2003xt,Gilkey:2004dm}. For Lifshitz operators these special techniques are not so powerful and we summarize them in App.~\ref{sec:func}.

\subsection{Factorization methods for product manifolds}\label{Fact}

In case the manifold $M=M_{1}\times M_{2}$ is a direct product of two manifolds $M_1$ and $M_{2}$ and the bundle over which the
operator $\D$ acts can be factorized correspondingly, such that the operator ${\cal D}$ can be decomposed in terms of two operators ${\cal D}_1$ and ${\cal D}_2$,
\be
\D=\D_1 \otimes \mathbb{I}_2+\mathbb{I}_1\otimes \D_2, \label{eq:facto_op}
\ee
which act trivially on one part of the direct sum of bundles, the trace (for a smearing function $F(x,y)=F_1(x) F_2(y)$ with $x$, $y$ coordinates on $M_{1}$ and $M_2$ respectively) of \eqref{eq:facto_op} simply reduces to the product of traces of the individual operators ${\cal D}_1$ and ${\cal D}_2$
\begin{align}
\label{eq:fact_tra}
\Tr_{M}\left(F_1 F_2 e^{-\tau \D}\right)=\Tr_{M_1}\left(F_1 e^{-\tau \D_1}\right)\Tr_{M_2}\left(F_2 e^{-\tau \D_2}\right).
\end{align}
Here the subscripts $M_{1}$ and $M_{2}$ for the functional traces indicate that the integration is performed only over the coordinates of $M_1$ and $M_{2}$, respectively.
If the order of $\D_1$ is $u$ and the order of $\D_2$ is $p\cdot u$, one can insert in \eqref{eq:fact_tra} the ansatz for the corresponding asymptotic expansions \cite{fegan1985,Gilkey:2004dm} and find that the coefficients $B_{n}[F,{\cal D}]$ of the total operator ${\cal D}$ is determined in terms of a sum of products of the coefficients $B_{i}[F_1,{\cal D}_{1}]$ and $B_{j}[F_2,{\cal D}_2]$ of the operators ${\cal D}_{1}$ and ${\cal D}_2$,
\be
\label{eq:factorization}
\AA_n[F,\D]=\sum_{p\,i+j=n} \AA_i[F_1,\D_1] \AA_j[F_2,\D_2].
\ee

\subsection{Heat kernel traces for powers of operators: Fegan-Gilkey lemma}\label{PO}

The heat kernel trace of a covariant spatial operator $\cal O(\nabla)$ of order $2u$, which acts on a bundle in a $\ds$-dimensional spatial manifold\footnote{Note that ${\cal O}(\nabla)$ is a {\it covariant} spatial operator which involves only covariant spatial derivatives $\nabla$.} has the asymptotic expansion
\be
\Tr \big(\,F e^{-\tau {\cal O}})=\sum_{k=0}^\infty B_k[F,{\cal O}\,\big] \tau^{(k-\ds)/2u},\label{eq:HKtrB}
\ee
with coefficients $B_{k}[F,{\cal O}]$.
Following \cite{fegan1985} (see also \cite{Gilkey:2004dm}), one can show that the asymptotic expansion for the trace of the heat kernel of the $n$-th power of $\cal O(\nabla)$, acted upon by the $m$-th power of ${\cal O}(\nabla)$ is given by,
\be
\label{eq:Omn}
\Tr\big(\,F\, {\cal O}^m e^{-\tau {\cal O}^n}\big)= \sum_{k=0}^\infty B_{k,n,m}[\,F,{\cal O}\,]\,{\tau^{(k-\ds-2um)/(2un)}},
\ee
with coefficients
\be\label{Gilkey_lemma}
B_{k,n,m}[\,F,{\cal O}\,]=
\lim_{\epsilon \to 0}\frac
{\Gamma\Big(\frac{\ds-k}{2u n}+\frac{m+\epsilon}{n}\Big)}{n\Gamma\Big(\frac{\ds-k}{2u}+\epsilon\Big)}\,
B_k[\,F,{\cal O}\,].
\ee
Notice that for $m, n\in \mathbb N$, the previous coefficient is always finite, because for a negative integer $(d-k+2 um)/(2nu)=-N$ the argument of the gamma function in the denominator is also negative integer $(d-k)/2u=-Nn-m$ (we  assume an integer value of $u$). Furthermore, it vanishes for  $(k-\ds)/2u\in \mathbb N$ unless  $(k-\ds-2um)/(2un)\in \mathbb N$. Thus, the asymptotic expansion \eqref{eq:Omn} with coefficients $B_{k,n,m}[F,{\cal O}]$ can be reduced to the calculation of the coefficients $B_{k}[F,{\cal O}]$ for the asymptotic expansion \eqref{eq:HKtrB} of the operator ${\cal O}(\nabla)$.

Note also the similarity of the above equations with our result \eqref{expansion_final}.
This is not surprising, because these equations can be obtained by expressing the operator ${\cal O}^m\exp[-\tau {\cal O}^n]$ as a contour integral involving the resolvent of ${\cal O}$, analogous to (\ref{Fourier}). As an illustration, both expressions coincide for the  limit $\varOmega=1$,  $u=1$ and $n=z$, $m=0$.

\subsection{Heat kernel traces for polynomials of operators}\label{RMAO}

For manifolds where  the factorization \eqref{eq:fact_tra} holds between space and time, the Fegan-Gilkey lemma can be  used to efficiently
compute the coefficients of the spatial operator if it  corresponds to a polynomial\footnote{One can always normalize the coefficient $c_n$ of the highest derivative part $(-\Delta)^n$ to one $c_n=1$, by factoring out a constant overall numerical factor $c_{n}$.}
\begin{align}
G(-\Delta)=(-\Delta)^n+\sum_{r=1}^{n-1} c_r (-\Delta)^r, \label{eq:PolynomOp}
\end{align}
with {\em constant} coefficients $\nabla_{i}c_r=0$. The heat kernel for this operator
 obeys the property,
\begin{align}
\exp{\left[G(-\Delta)\right]}=\exp{\left[\sum_{r=0}^{n}c_r\,(-\Delta)^r\right]}=
\prod_{r=0}^{n}\,\exp{\left[c_r\,(-\Delta)^r\right]},\label{eq:SepHK}
\end{align}
which allows one to write the heat kernel trace in the form
\begin{align}
 \Tr \left( F e^{-\tau G(-\Delta)}\right) ={}&\Tr \left( F e^{-\tau \left(\sum_{r=1}^{n-1} c_r (-\Delta)^r\right)}\;e^{-\tau(-\Delta)^n}\right),
\end{align}
Expanding the first exponential in powers of $\sum_{r=1}^{n-1} c_r (-\Delta)^r$ and using the linearity of the trace, we obtain for the heat kernel trace,
    \be
    \Tr ( F e^{-\tau G(-\Delta)})=\sum_{k=0}^\infty c_{n,k}\,\Tr \left( F\; (-\Delta)^k\,e^{-\tau (-\Delta)^n}\,\right),\label{eq:TrGb}
    \ee
with some new constant coefficients $c_{n,k}$ that can be easily retrieved.
In this form, the reduction method \eqref{eq:Omn}, \eqref{Gilkey_lemma} can directly be applied, such that the asymptotic expansion of \eqref{eq:TrGb} can ultimately be expressed in terms of the heat kernel coefficients of the covariant Laplacian $B_{k}[F,-\Delta]$. In section \ref{sec:examples}, we illustrate how this technique can be used to calculate the critical coefficient $\AA_{d+z}[F,{\cal D}]$ for an {\it anisotropic} operator, whose spatial part is separable.  The connection to \eqref{expansion_final} is also clear and one can easily check that both expressions agree when applied to the same family of operators.

\section{Effective action for  Lifshitz theories in projectable manifolds}\label{sec:calcul}

In this section we focus on the calculation of the local heat kernel coefficient $b_{d+z}[\D]$ for {\emph{projectable}} backgrounds, i.e. metrics   for which the acceleration vector $a_i$ vanishes or, equivalently, the lapse function $N$ only depends on time (cf. \eqref{eq:Ka}). The projectability condition is compatible with FDiff transformations \eqref{eq:FDiff} \cite{Horava:2009uw} and it is especially interesting in the case of Ho\v rava gravity, where the projectable theory has been shown to be renormalizable \cite{Barvinsky:2015kil}.

\subsection{General results}

We work in the gauge $N=1$, $N^i=0$, which, as explained in section \ref{sec:curved}, is always accessible if the lapse only depends on time. Moreover, we assume that the dynamical field $\phi^A$ is an $s$-tuple of scalar fields, and that the bundle metric $G_{AB}$ is a constant matrix. Then the operator (\ref{eq:gen_D}) (see also \eqref{eq:oper_form}) reads in anisotropic units
\begin{eqnarray}
&&\D=-\frac{1}{\sqrt\gamma}\,\partial_t\sqrt\gamma\,\partial_t+{\cal O},\label{eq:op_proj}\\
&&{\cal O}=(-\Delta)^z+\sum_{m=0}^{2z-1}\Omega_{(m)}^{i_1\dots i_m}\nabla_{i_1}\dots\nabla_{i_m}.\label{eq:O}
\end{eqnarray}
Recall that the fields $\Omega_{(m)}$ carry extra indices $A$ and $B$, which in the present context means that they are matrix-valued tensor fields. Our strategy for computing the coefficient $b_{d+z}[\D]$ is the following: first we use Eqs.~(\ref{eq:Tconstraint}) and (\ref{eq:dim_term}), which are based on time-reversal invariance and dimensional analysis, to determine the structure of this coefficient; then we use that information to map the problem to a simpler problem, namely the computation of the same coefficient on a more specific background; and finally, if necessary, we apply the resolvent method described in section \ref{sec:general}. We will consider the cases $z>d$ and $z=d$ separately. We do not study the case $z<d$, mainly because gravitational Lifshitz theories with $z<d$ are power-counting non-renormalizable.

Before starting, it is convenient to take a closer look at the space on which $\D$ acts. This is the space ${\cal F}({\cal M},{\mathbb R}^s)$ of functions ${\cal M}\to{\mathbb R}^s$, where ${\cal M}$ denotes the space-time manifold. Let us further assume that the space-time manifold has the form ${\cal M}=\Sigma\times{\mathbb R}$, so that $t$ is the standard coordinate on ${\mathbb R}$ and $x^i$ are coordinates on $\Sigma$ (note that there is no loss of generality in this assumption, because our aim is to compute a local quantity which is not sensitive to the topological properties of space-time).
Then we have a  tensor product decomposition
\begin{equation}\label{eq:tensor2}
{\cal F}({\cal M},{\mathbb R}^s)={\cal F}({\mathbb R})\otimes{\cal F}(\Sigma,{\mathbb R}^s)
\end{equation}
The first factor of this decomposition can be thought of as ``functions of time'', and the second as ``multiplets of functions of space''. Since the background fields, $\gamma_{ij}$ and $\Omega_{(m)}$, depend on both time and space, in general none of the terms in (\ref{eq:op_proj}) and (\ref{eq:O}) act trivially on any of the factors of this decomposition.  This is however the case  for particular backgrounds  that can be used to find the general result, as we will see momentarily.

\subsubsection*{The case $z>d$}
Let us start with the case $z>d$. Setting $n=d+z$ in (\ref{eq:dim_term}) and using (\ref{eq:Tconstraint}) we see that, in this case, $b_{d+z}[\D]$ involves \emph{no time derivatives of the background fields}, $\gamma_{ij}$ and $\Omega_{(m)}$. Therefore, it is sufficient to consider a background manifold with vanishing extrinsic curvature and no time dependence at all in order to capture the whole structure in $b_{d+z}[\D]$. For such background, the operator \eqref{eq:op_proj} reduces to
\begin{equation}\label{eq:barD}
\bar\D=-\partial_{t}^2+{\cal O}.
\end{equation}
On this background, this operator acts on \eqref{eq:tensor2} as
\begin{equation}\label{eq:fact_op}
\bar\D=-\partial_{t}^2\otimes{\mathbb I}+{\mathbb I}\otimes{\cal O}
\end{equation}
where ${\mathbb I}$ denotes the identity. Therefore, the heat operator factorizes as described in
\eqref{eq:fact_tra}.
The first trace of this factorization is easily computed
\begin{equation}\label{eq:trace1}
\Tr\left[F_1\,e^{-\tau(-\partial_{t}^2)}\right]=\frac{1}{\sqrt{4\pi\tau}}\int\di t\,F_1,
\end{equation}
for functions $F_1$ with appropriate boundary conditions.
On the other hand, an argument analogous to the one we used in section \ref{sec:eff_heat} shows that the second trace admits the asymptotic expansion
\begin{equation}\label{eq:trace2}
\Tr\left(F_2\,e^{-\tau{\cal O}}\right)=\frac{1}{\tau^{d/2z}}\sum_{n=0}^\infty \int\di^d x\,\sqrt{\gamma}\,F_2\,b_n[{\cal O}]\,\tau^{n/2z},
\end{equation}
with $F_2\in{\cal F}(\Sigma)$.
Here, $b_n[{\cal O}]$ is a linear combination of scalars constructed out of the spatial Riemann tensor $R_{ijkl}$, the fields $\Omega_{(m)}$ and their spatial derivatives, with coefficients that only depend on $d$ and $z$, and which vanishes for odd $n$.

Substituting (\ref{eq:trace1}) and (\ref{eq:trace2}) into \eqref{eq:factorization} (e.g. comparing with the asymptotic expansion  of the heat kernel for the trace of $\bar\D$ (\ref{eq:asymp})) we obtain
\begin{equation}\label{eq:bfact}
b_n[\bar\D](t,\mathbf{x})=\frac{1}{\sqrt{4\pi}}b_n[{\cal O}](\mathbf{x}).
\end{equation}
Finally, from this equation and the covariant structure of the result, we can forget about the barred operator and generalize this result to any operator of the form \eqref{eq:op_proj}-\eqref{eq:O}, namely
\begin{equation}\label{eq:z>d}
b_{d+z}[\D]=\frac{1}{\sqrt{4\pi}}b_{d+z}[{\cal O}]\qquad{\text{for }}z>d,
\end{equation}
where $b_{d+z}[{\cal O}](t,\mathbf{x})$ is the heat kernel coefficient of a standard elliptic operator, which has been thoroughly studied in the literature (cf. \cite{Vassilevich:2003xt,Barvinsky:1985an,kirstenbook,Gilkey:2004dm} and references therein). As an important consequence of the preceding analysis, we therefore arrive at the statement that \emph{the heat kernel critical coefficient of a projectable Lifshitz operator in the case $z>d$ is given by the coefficient in the asymptotic expansion of the purely spatial part of the operator.}

\subsubsection*{The case $z=d$}
Next we turn our attention to the case $z=d$. Setting $n=d+z$ in (\ref{eq:dim_term}) and using (\ref{eq:Tconstraint}) we find that, in this case, the coefficient of interest has the form
\begin{equation}\label{eq:bTV}
b_{d+z}[\D]=T[\D]+V[\D],
\end{equation}
where $T[\D]$ (the ``kinetic term'') is a linear combination of $K_{ij}K^{ij}$, $K^2$ and $D_tK$, and $V[\D]$ (the ``potential term'') is a linear combination of scalars constructed out of the spatial Riemann tensor $R_{ijkl}$, the fields $\Omega_{(m)}$ and their spatial derivatives. The coefficients of both linear combinations only depend on $d$.

Since the potential term $V[\D]$ involves no time derivatives of the background fields and the kinetic term $T[\D]$ vanishes if the background fields are time-independent, we clearly have a situation analogous to the previous case. If we choose a set of background fields with no time dependence, we can repeat the arguments of the case $z>d$ and find that
\begin{equation}
V[\D]=\frac{1}{\sqrt{4\pi}}b_{d+z}[{\cal O}].\label{eq:V}
\end{equation}

Things are not so easy  for the kinetic term $T[\D]$.  In particular, it is not sufficient to go to a  background with no spatial dependence because even in that situation, the two parts \eqref{eq:op_proj} and \eqref{eq:O} of the operator will not commute and therefore we cannot split the traces.
However, since we have already captured all the dependence in spatial derivatives, we can use a simplified form
with a purely time-dependent metric\footnote{This is essentially equivalent to consider the generic metric evaluated at one spatial point,
and disregarding the effect of spatial derivatives.} $\gamma(t)^{ij}\equiv \gamma_t^{ij}$
\begin{equation}\label{eq:D'}
\D'=-\frac{1}{\sqrt{\gamma_t}}\,\partial_t\sqrt{\gamma_t}\,\partial_t+(-\Delta_{\gamma_t})^z.
\end{equation}
Now we can use the resolvent method described in section \ref{sec:general}. Before doing that, it is convenient to make an important observation. Eqs.~(\ref{eq:Tconstraint}) and (\ref{eq:dim_term}) imply that $b_n[\D']$ is non-vanishing only when $n/z$ is an even integer, and in that case it involves precisely $n/z$ derivatives of the spatial metric. From this we see that, in order to determine the coefficient $b_{d+z}[\D']$ in the case $z=d$, it suffices to compute the heat kernel trace of (\ref{eq:D'}) up to two-derivative order. This was done, using the resolvent method, in appendix C of \cite{Nesterov:2010yi}, setting the smearing function $F$ to one and hence neglecting total derivatives. From the results of that reference, restoring total derivatives, we find for arbitrary $d$ and $z$
\begin{alignat}{2}\label{eq:spaceindep}
b_{d+z}[\D'](t,\mathbf{x})&=-\frac{1}{(4\pi)^{\frac{d+1}{2}}}\frac{\Gamma\left(\frac{d}{2z}\right)}{\Gamma\left(\frac{d}{2}\right)}\times \;\nonumber\\
&\text{tr}\bigg[\frac{\mathbb{I}}{2(d+2)}\left(\frac{d+2z}{3z}K_{ij} K^{ij}-\frac{d+3-z}{3z} K^2\right)+\frac{\mathbb{I}}{3z}(D_t+ K) K\bigg](t).
\end{alignat}
where the trace and the identity $\mathbb{I}$ are evaluated over the bundle indices $A,B$.

The last term in \eqref{eq:spaceindep} is the total derivative that was dropped in \cite{Nesterov:2010yi}. Setting $z=d$ in this equation and restoring the full space dependence of the background fields, we obtain the kinetic term,
\begin{equation}\label{eq:T}
T[\D]=-\frac{s}{(4\pi)^{\frac{d+1}{2}}}\frac{\sqrt{\pi}}{\Gamma\left(\frac{d}{2}\right)}\,\bigg[\frac{1}{2(d+2)}\left( K_{ij} K^{ij}-\frac{1}{d} K^2\right)
+\frac{1}{3d}(D_t+K) K\bigg].
\end{equation}
where we have explicitly written $\text{tr}\,\mathbb{I}=s$ for the case of the s-tuple of scalar fields at hand.

Finally, substituting the expressions (\ref{eq:T}) and (\ref{eq:V}) for the kinetic and potential terms into (\ref{eq:bTV}) and generalizing to the full operator we arrive at the result
\begin{eqnarray}\label{eq:z=d}
b_{2d}[\D]&=&\frac{1}{\sqrt{4\pi}}\; b_{2d}[{\cal O}]\nonumber\\
&&-\frac{s}{(4\pi)^{\frac{d+1}{2}}}\frac{\sqrt{\pi}}{\Gamma\left(\frac{d}{2}\right)}\bigg[\frac{1}{2(d+2)}\left( K_{ij} K^{ij}-\frac{1}{d} K^2\right)
+\frac{1}{3d}(D_t+K) K\bigg].\end{eqnarray}

Again we find that \emph{the heat kernel critical coefficient of a projectable Lifshitz operator in the case $z=d$ is given by the coefficient in the asymptotic expansion of the purely spatial part of the operator plus a universal term containing only extrinsic curvatures.}

Eqs.~(\ref{eq:z>d}) and (\ref{eq:z=d}) are the main results of this section. They express the heat kernel coefficient of interest in terms of a heat kernel coefficient of an elliptic operator, which is known in several cases. In the next section we explore some of these cases, thus obtaining explicit results for the effective action of projectable Lifshitz theories.

\subsection{Particular cases}\label{sec:examples}

We now apply the previous methods to find the one-loop effective action for some particular models. As described in section \ref{sec:intro}, the 
landscape of Lifshitz models (even for the projectable case!) is rather large and a systematic study is beyond the purpose of this paper.
Instead, we  focus on models that illustrate the main advantages of our methods and that can serve as inspiration for further studies. In other words,  we prefer to illustrate our computational tools with theories for which the calculation can be done efficiently and that are relatively  
generic, so that applications to physical systems may be found. 

We will consider two main cases, namely the case $z=d=2$ with $\Omega_{(m)}$ (almost) arbitrary, and the case where $d$ and $z$ are arbitrary and the fields $\Omega_{(m)}$ are such that the operator (\ref{eq:O}) is a polynomial in the Laplacian with coefficients that only depend on time. Some partial results can be checked with the functorial properties of appendix~\ref{sec:func}, though we will not be explicit about this possibility.

\subsubsection*{Case $z=d=2$}
In the case $z=d=2$, the operator (\ref{eq:O}) involves four coefficient fields $\Omega_{(m)}$, corresponding to $m=0,1,2,3$. Suppose that the last of these fields, $\Omega_{(3)}$, vanishes\footnote{For the general case  $\Omega_{(3)}\neq0$, the global  quantity $B_4[1,{\cal O}]$ can be found in \cite{Barvinsky:1985an}.}. Then the heat kernel coefficient $b_4[{\cal O}]$ is known
  \cite{Fradkin:1981hx,Barvinsky:1985an,Gusynin1990296},
\begin{alignat}{2}
b_4[{\cal O}]=\frac{1}{16\sqrt\pi}\,\mathrm{tr}\,\bigg[\frac{1}{4}\Delta \Omega_{(2)} - \frac{1}{2}\nabla_i\nabla_j \Omega^{ij}_{(2)} +
\frac{1}{16}\,\Omega^{2}_{(2)} + \frac{1}{8}\,\Omega^{ij}_{(2)}\Omega^{\phantom{ij}}_{(2)ij}&\nonumber\\
+ \frac{1}{6}\,\Omega_{(2)} R - \frac{1}{3}\,\Omega^{ij}_{(2)}R_{ij} +\nabla_i \Omega^{i}_{(1)}-2\,\Omega_{(0)}&\bigg],
\label{eq:gusy}
\end{alignat}
where $\tr\, C\equiv C^{A}_A$ and $\Omega_{(2)}\equiv\gamma_{ij}\Omega_{(2)}^{ij}$, and where it is assumed that $\Omega_{(2)}^{ij}$ is symmetric, which implies no loss of generality because two covariant derivatives commute when acting on a scalar field (see (\ref{eq:O})). We have independently reproduced this result using the covariant analog of the method developed in section \ref{sec:general}. From (\ref{eq:z=d}) and (\ref{eq:gusy}) we obtain
\begin{alignat}{2}\label{eq:z=d=2}
b_4[\D]=&-\frac{s}{16\pi}\,\bigg[\frac{1}{4}\left( K_{ij} K^{ij}-\frac{1}{2} K^2\right)
+\frac{1}{3}(D_t+K) K\bigg]&&\nonumber\\
&+\frac{1}{32\pi}\,\mathrm{tr}\,\bigg[\frac{1}{4}\Delta \Omega_{(2)} - \frac{1}{2}\nabla_i\nabla_j \Omega^{ij}_{(2)} +
\frac{1}{16}\,\Omega^{2}_{(2)} + \frac{1}{8}\,\Omega^{ij}_{(2)}\Omega^{\phantom{ij}}_{(2)ij}\nonumber\\
&\ \ \ \ \ \ \ \ \ \ \ \ \ \ \ \ \ \ \ \ \ + \frac{1}{6}\,\Omega_{(2)} R - \frac{1}{3}\,\Omega^{ij}_{(2)}R_{ij} +\nabla_i \Omega^{i}_{(1)}-2\,\Omega_{(0)}\bigg].
\end{alignat}
This coefficient was previously computed (with no restrictions on the lapse function) in \cite{Baggio:2011ha} for $\Omega_{(m)}=0$ and in \cite{Griffin:2012qx} for $\Omega_{(2)}=\Omega_{(1)}=0$, $\Omega_{(0)}\ne 0$. Both results agree with (\ref{eq:z=d=2}) in the corresponding limits.

\subsubsection*{Polynomials of the Laplacian: general formulae}

Let us now return to generic values of $d$ and $z$. Suppose that $\Omega_{(m)}$ vanishes for odd $m$, and that, for each $r=0,\dots,z-1$, there exists a matrix-valued scalar field $\hat{\alpha}_r$ such that
\begin{equation}
\Omega_{(2r)}^{i_1\dots i_{2r}}=(-1)^r\,\hat{\alpha}_r\,\gamma^{i_1i_2}\dots\gamma^{i_{2r-1}i_{2r}}.
\end{equation}
Then the spatial part of the operator, given by (\ref{eq:O}), is a polynomial in the Laplacian,
\begin{equation}
{\cal O}=(-\Delta)^z+\sum_{r=0}^{z-1}\hat{\alpha}_r(-\Delta)^r.
\end{equation}
Suppose also that $\hat{\alpha}_r$ only depends on time, $\hat{\alpha}_r(t,\mathbf{x})=\hat{\alpha}_r(t)$. Then the first term above commutes with the remaining terms and, in consequence, the heat operator of ${\cal O}$ is equal to the heat operator of the first term times the heat operator of the sum of the remaining terms. Expanding the latter in Taylor series we obtain
\begin{equation}
e^{-\tau{\cal O}}=\sum_{k=0}^\infty\frac{(-\tau)^k}{k!}\sum_{r_1=0}^{z-1}\dots\sum_{r_k=0}^{z-1}\hat{\alpha}_{r_1}\dots \hat{\alpha}_{r_k}(-\Delta)^{r_1+\dots+r_k}e^{-\tau(-\Delta)^z}.
\end{equation}
Recall that, at each time $t$, we are viewing ${\cal O}$ as an operator on the space ${\cal F}(\Sigma,{\mathbb R}^s)$ of multiplets of functions of space, which admits the decomposition ${\cal F}(\Sigma,{\mathbb R}^s)={\cal F}(\Sigma)\otimes{\mathbb R}^s$. If we think of $\Delta$ and $\hat{\alpha}_r$ as operators on the first and second factors of this decomposition respectively, the above product between a function of the matrices $\hat{\alpha}_r$ and a function of $\Delta$ takes the form of a tensor product, so we have
\begin{equation}\label{eq:tracepol}
\Tr\left(Fe^{-\tau{\cal O}}\right)=\sum_{k=0}^\infty\frac{(-\tau)^k}{k!}\sum_{r_1=0}^{z-1}\dots\sum_{r_k=0}^{z-1}\tr\,(\hat{\alpha}_{r_1}\dots \hat{\alpha}_{r_k})\,\Tr\left[F(-\Delta)^{r_1+\dots+r_k}e^{-\tau(-\Delta)^z}\right],
\end{equation}
where, again, $\tr\, C\equiv C^{A}_A$ and, since $\Delta$ is viewed as an operator on ${\cal F}(\Sigma)$ (i.e., as acting on a single scalar field), the corresponding trace has to be understood accordingly.
Now we can use the result \eqref{Gilkey_lemma} to compute the heat trace of the operator above. Comparing with the asymptotic expansion (\ref{eq:trace2}) for the heat kernel trace of ${\cal O}$ and concentrating on the coefficient of interest for us we obtain
\begin{eqnarray}
&&b_{d+z}[{\cal O}]=\sum_{n=0}^{d+z} c_nb_n[-\Delta],\label{eq:bpol}\\
&&c_n=\frac{1}{z\Gamma\left(\frac{d-n}{2}\right)}\sum_k\frac{(-1)^k}{k!}\Gamma\left(k-\frac{1}{2}\right)\sum_{r_1,\dots,r_k}\tr\,(\hat{\alpha}_{r_1}\dots\hat{\alpha}_{r_k}),\label{eq:cn}
\end{eqnarray}
where the sums in the second equation run over those values of $k\in{\mathbb N}$, $r_i\in\{0,\dots,z-1\}$ which satisfy the constraints
\begin{equation}\label{eq:constraints}
\frac{d+z-n}{2z}\le k\le \frac{d+z-n}{2},\qquad  r_1+\dots+r_k=zk-\frac{d+z-n}{2}.
\end{equation}
Eqs.~(\ref{eq:bpol})-(\ref{eq:constraints}) give the heat kernel coefficient $b_{d+z}[{\cal O}]$ in terms of the first $d+z+1$ heat kernel coefficients of the Laplacian acting on a single scalar field, $b_n[-\Delta]$. Of course, the latter vanish for odd $n$, and explicit formulae for even $n$ can be found in the literature up to $n=10$ (see \cite{Vassilevich:2003xt} and references therein).
As a check of the above result, note that, for $n$ even, the second equation in (\ref{eq:constraints}) can never be satisfied if $d+z$ is odd, so $b_{d+z}[{\cal O}]=0$ for odd $d+z$, as it should be. We have also checked that the above result agrees with (\ref{eq:gusy}) in the case $z=d=2$. Let us now explore two more subcases.

\subsubsection*{Polynomials of the Laplacian with $z=d=3$}
Setting $z=d=3$ in (\ref{eq:bpol})-(\ref{eq:constraints}) we obtain
\begin{eqnarray}\label{eq:bpol3}
b_6[{\cal O}]&=&-\frac{1}{2}\,b_6[-\Delta]\,s+\frac{1}{6}\,b_4[-\Delta]\,\tr\,\hat{\alpha}_2 +\frac{1}{3}\,b_2[-\Delta]\,\tr\left(-\hat{\alpha}_1+\frac{1}{4}\hat{\alpha}_{2}^2\right)\nonumber\\
&&~~~~~~~~~~~~~~~~~~~~~~~~~~~~~~~~+\frac{2}{3}\,b_0[-\Delta]\,\tr\left(-\hat{\alpha}_0+\frac{1}{2}\hat{\alpha}_1\hat{\alpha}_2-\frac{1}{8}\hat{\alpha}_{2}^3\right),
\end{eqnarray}
which, together with (\ref{eq:z=d}), implies
\begin{eqnarray}\label{eq:z=d=3}
b_{6}[\D]&=&-\frac{s}{8\pi^2}\,\bigg[\frac{1}{10}\left( K_{ij} K^{ij}-\frac{1}{3} K^2\right)
+\frac{1}{9}(D_t+K) K\bigg]\nonumber\\
&&+\frac{1}{\sqrt{4\pi}}\,\bigg\{-\frac{1}{2}\,b_6[-\Delta]\,s+\frac{1}{6}\,b_4[-\Delta]\,\tr\,\hat{\alpha}_2 +\frac{1}{3}\,b_2[-\Delta]\,\tr\left(-\hat{\alpha}_1+\frac{1}{4}\hat{\alpha}_{2}^2\right)\nonumber\\
&&\ \ \ \ \ \ \ \ \ \ \ \ \,+\frac{2}{3}\,b_0[-\Delta]\,\tr\left(-\hat{\alpha}_0+\frac{1}{2}\hat{\alpha}_1\hat{\alpha}_2-\frac{1}{8}\hat{\alpha}_{2}^3\right)\bigg\}.
\end{eqnarray}
For the convenience of the reader, we have listed the heat kernel coefficients of the Laplacian (evaluated as acting on a single scalar field) which appear in these equations in appendix \ref{HK_laplace}.  In the case at hand of a single scalar field $\phi^A=\varphi$, the bundle curvature vanishes ${\cal R}_{\mu\nu}=0$ and $\tr\,\mathbb{I}=1$.

\subsubsection*{Case $d$ and $z$ arbitrary, $\Omega_{(m)}=0$}

Note that this corresponds to setting $\hat{\alpha}_r=0$ in (\ref{eq:cn}).  We find that the only term that contributes to $c_n$ is the term $k=0$, which, by the first equation in (\ref{eq:constraints}), is present only when $n=d+z$. Thus, (\ref{eq:bpol}) reduces to
\begin{equation}\label{eq:balpha0}
b_{d+z}[{\cal O}]=-\frac{\sqrt{4\pi}}{z\Gamma\left(-\frac{z}{2}\right)}\,s\,b_{d+z}[-\Delta].
\end{equation}
This result can also be obtained directly from \eqref{Gilkey_lemma} by noting that in the case $\hat{\alpha}_r=0$, $B_{d+z}[F,{\cal O}]=B_{d+z,0,z}[F,-\Delta]$  (with $\Delta$ viewed as acting on a multiplet of scalar fields). Substituting into (\ref{eq:z>d}) and (\ref{eq:z=d}) we obtain
\begin{eqnarray}
b_{d+z}[\D]&=&-\frac{s}{z\Gamma\left(-\frac{z}{2}\right)}\,b_{d+z}[-\Delta]\qquad{\text{for }}z>d,\label{eq:zMdO0}\\
b_{d+z}[\D]&=&-\frac{s}{(4\pi)^{\frac{d+1}{2}}}\frac{\sqrt{\pi}}{\Gamma\left(\frac{d}{2}\right)}\,\bigg[\frac{1}{2(d+2)}\left( K_{ij} K^{ij}-\frac{1}{d} K^2\right)
+\frac{1}{3d}(D_t+K) K\bigg]\nonumber\\
&&~~~~~~~~~~~-\frac{s}{z\Gamma\left(-\frac{z}{2}\right)}\,b_{d+z}[-\Delta],\qquad {\text{for }}z=d.\label{eq:zdO0}
\end{eqnarray}
Note that the right-hand side of  \eqref{eq:zMdO0} vanishes when either $z$ is even or $d+z$ is odd. In other words, it vanishes unless both $d$ and $z$ are odd. We may thus conclude that {\emph{projectable Lifshitz theories with $z>d$ and $\Omega_{(m)}=0$ are one-loop finite unless both $d$ and $z$ are odd}}.

\section{Conclusion and future work}\label{sec:concl}

Quantum Lifshitz theories are currently being explored in a variety of fields, from condensed matter \cite{Fradkin:1991nr,Ardonne:2003wa,Parker:2017lnh} to quantum gravity  \cite{Horava:2009uw,Barvinsky:2015kil}. This interest stems from the adequacy of these theories to describe the behaviour close to fixed points with anisotropic properties between space and time, which may be present in different physical situations or even at a fundamental level.  In this work we have developed a formalism to
efficiently compute the one-loop effective action for such theories in curved backgrounds. We have done it by the heat kernel techniques that relate the effective action to the evaluation of a concrete coefficient of the asymptotic expansion of the trace of the heat kernel, cf.  \eqref{eq:effact} and \eqref{eq:asymp}.

Our strategy to compute $B_{d+z}[F,{\cal D}]$ for $\cal D$ defined in \eqref{eq:gen_D} is divided into two possibilities: first, in section~\ref{sec:general} we have closely followed  \cite{Nesterov:2010yi} and provided a formalism for general operators based on a resolvent method that allows one to write the asymptotic expansion \eqref{eq:asymp} in terms of the expansion of a Lorentz invariant operator, equation \eqref{expansion_final_3}. We have carefully defined the method and provided a rigorous justification of all the steps (see also \cite{D'Odorico:2015yaa} for a related technique). Still, this strategy is quite cumbersome in general.  This can be overcome for certain operators, which is the basis of our second suggestion: to use Fegan-Gilkey lemma \eqref{eq:Omn} and other simplification techniques of the general formalism presented in section~\ref{sec:calc_fo} (see also Appendix \ref{sec:func}). This is particularly relevant for the case of {\it projectable} manifolds, which is the subject of section \
\ref{sec:calcul}. The latter section presents a series of original results based on the application of our simplified techniques to
compute the effective action for operators of the form  \eqref{eq:op_proj}. The main generic results are collected in the equations \eqref{eq:z>d} and \eqref{eq:z=d}.
We have also applied these methods to some particular operators, with the outcome  that can be read from  equations \eqref{eq:z=d=2}, (\ref{eq:bpol})-(\ref{eq:constraints}), \eqref{eq:z=d=3} and \eqref{eq:zMdO0}-\eqref{eq:zdO0}. The aim of this exercise is to show the efficiency  of the whole approach with some concrete examples.
Independently of this method, we have used scaling arguments to generalize the condition of one-loop finiteness of relativistic theories in odd space-times to the Lifshitz case, where it happens for the cases with  $\ds+z$  odd.

The formalism that we have developed opens the door to a plethora of applications and further investigations.
A first concrete example is the calculation of the effective action of Ho\v rava gravity, as already started in \cite{D'Odorico:2015yaa,Griffin:2017wvh}.  This is a very  important calculation,
necessary to understand the consistency of the proposal as a complete theory of quantum gravity.
For generic Lifshitz theories, one can try to expand previous works on Weyl anomalies  \cite{Baggio:2011ha,Adam:2009gq,Griffin:2011xs,Gomes:2011di,Arav:2014goa,Pal:2017ntk} and perform more explicit calculations. This may be relevant for  the possible classification of renormalization group flows, as happens in the relativistic case  \cite{Zamolodchikov:1986gt,Komargodski:2011vj,Nakayama:2013is}, see e.g. \cite{Nakayama:2012sn} for some first attempts. Furthermore, the classification and calculation of anomalies is important to understand the possible gravity duals of non-relativistic theories \cite{Baggio:2011ha,Adam:2009gq,Griffin:2011xs}.  Finally, the previous results may be extended to conical geometries  and used to  establish the properties about entanglement entropy in Lifshitz theories via the replica trick \cite{Callan:1994py}, see  \cite{Nesterov:2010yi}  for some first results in this direction.  The results on entanglement entropy may also be important to study the gravitational duals of 
Lifshitz theories, as happens in the relativistic case \cite{Ryu:2006bv,Emparan:2006ni}.
As we discussed in the introduction, all these lines of research would help to better understand generic properties of  quantum field theory beyond the relativistic case.

Finally, the methods we have presented can also be applied for  covariant operators that can be expressed as functions of second order operators, as should be clear by considering the purely spatial part of \eqref{eq:gen_D}.  In particular, this is evident from results such as the Fegan-Gilkey lemma \eqref{Gilkey_lemma}. Thus, the techniques we introduced and extensions thereof may be relevant to investigate the generic properties of these relativistic (not necessarily local) quantum field theories, including aspects of holography, entanglement entropy or RG flows, see e.g. \cite{Shiba:2013jja,Camps:2013zua}.


\section*{Acknowledgements}

We are indebted to Sergey Sibiryakov, Enrique \'Alvarez, Andreas Stergiou, Carlos Hoyos and Solomon Endlich for stimulating discussions. A.B. is grateful for hospitality of CERN where this work was completed. His work was also supported in part by the RFBR grant No. 17-02-00651 and by the Tomsk State University Competitiveness Improvement Program. M.H-V is grateful for the hospitality of the theory department at CERN. He also wants to thank the Instituto de Física Teórica UAM/CSIC, whose warm and stimulating environment hosted part of this work. M.H-V has been supported by the Tomalla Foundation, European Union FP7 ITN INVISIBLES (Marie Curie Actions, PITN-GA-2011-289442), COST action MP1210 (The String Theory Universe) and by the Spanish MINECO Centro de Excelencia Severo Ochoa Programme under grant SEV-2012-0249. C.S. is grateful for the hospitality of the CERN theory division, where part of this work has been done. GPN acknowledges support from Fundación Bunge y Born, UBA and CONICET.
\appendix

\section{Evaluation of the asymptotic expansion in the generic case}\label{app:1}
\subsection{Scaling dependence of invariants and efficiency of  curvature expansion}\label{app:andrei}
We want to show that the series in $\tau$ \eqref{expansion_final} starts at non-negative powers and that it involves only a finite number of invariant terms at each polynomial order.
As we saw in section \ref{sec:AGB}, the curvature invariants with the anisotropically rescaled metric (\ref{tilde_g}) participating in the construction of the heat kernel of the operator \eqref{tildekernel}  have the form of typical Schwinger-DeWitt coefficients with the space-time metric $\tilde g_{\mu\nu}$ (recall \eqref{kernel}). In the particular ADM $(\ds+1)$-decomposition and by using \eqref{eq:ADM_t}, these curvature invariants  can be rewritten in terms of the spatial ($\ds$-dimensional) curvature $\tilde R$, spatial covariant derivatives $\tilde \nabla$, extrinsic curvature of spatial slices $\tilde K$ and time reparametrization invariant derivatives $\tilde D_t$. In the FDiff covariant formulation only $\tilde D_t$ and $\tilde K$ involve a factor of $\tilde N^{-1}$, such that the rescaling of the lapse function \eqref{eq:ADM_t} implies,
   \begin{eqnarray}
   \tilde D_t=\left(\frac\tau\lambda\right)^{1/2}D_t,\quad
   \tilde K=\left(\frac\tau\lambda\right)^{1/2}K. \label{eq:K_sca}
   \end{eqnarray}
Thus, bearing in mind Gauss-Codazzi relations between space-time curvature and spatial curvature, $R^{(\ds+1)}\sim R+K^2$, and a similar relation between space-time derivatives and spatial derivatives $D\sim\nabla+K +a$ \cite{Kuchar:1976yx}, the operator $e^{-\lambda \tilde {\cal A}}$ in the trace \eqref{structure} would give contributions that in terms of the original metric coefficients can be rewritten as a sum of  variously weighted  structures
   \begin{eqnarray}
 \frac{1}{\lambda^{(d+1)/2}}     \left(\frac\lambda\tau\right)^{1/2}
 \sum\limits_{v=0}^u \mathfrak{B}_{2u-2v,2v}
   \lambda^{u-v}\tau^v,                               \label{A_scaling}
    \end{eqnarray}
where
   \begin{eqnarray}
   &&\mathfrak{B}_{2u,2v}=\sum\limits_{l,p,r,s,q,s_{m}}\stackrel{r+s=2v,\,\,2l+p+q=2u}
   {\overbrace{D_t^r\,\nabla^p\,
R^l\,a^q \,K^s}},              \label{R}
   \end{eqnarray}
with clear notation.

An important observation regarding the sum (\ref{A_scaling}) is that after $\lambda$ and $q$ integrations in \eqref{expansion1} all terms with $v>0$ are of higher power in $\tau$.  Notice first that  integration over powers of the parameter $\lambda$  generates the power of
   \begin{eqnarray}
   \lambda\sim\frac1q\sim
   \frac{\big(M^2\tau\big)^{1/z}}{M^2}.             \label{q_to_tau}
   \end{eqnarray}
As a result,
    \begin{eqnarray}
   \frac\tau\lambda\sim
    (M^2\tau)^{1-1/z}
   \end{eqnarray}
with $z>1$. Therefore, if in what follows we would be able to prove the efficiency of the curvature expansion for the contribution of these $v=0$ terms then this proof would automatically apply to the rest of the terms of higher power in $\tau$.

To include the effect the Zassenhaus operator $\varOmega$, note it generates an infinite chain of derivatives  acting on the exponentiated world function as in (\ref{nabla_exp}). This gives negative powers of $\lambda$. To analyze them, let us first consider the case of ${\cal C}=0$ in (\ref{total_Zassenhaus}) -- the effect of nonzero ${\cal C}$ will be considered later. The structure of $\varOmega$ expanded in derivatives (their commutators form the coefficients of these derivatives) is as follows
   \begin{eqnarray}
   &&\varOmega=\sum\limits_{k,r,s,l,p\geq 0}
   \mathfrak{B}_k\frac{\tau^r\,
   \lambda^s}{M^{2p(z-1)}}D^l,                      \label{Omega1}\\
   &&2r+2s+2p(z-1)=l+k,                             \label{range1}
   \end{eqnarray}
where $\mathfrak{B}_k$ denotes a local quantity  of a positive\footnote{A negative dimensionality of $\mathfrak{B}_k$ would be  possible if we included the fields $\Omega_{(m)}$ in 
our analysis. This is the reason for first working with ${\cal C}=0$ and consider the effect of $\Omega_{(m)}\neq0$ later on. } physical dimensionality $k$, which is built similarly to (\ref{R}) entirely of curvatures and independent of $M$. Summation runs over indices satisfying the equality above, which guarantees the dimensionless nature of $\varOmega$. Powers $r,s,p,k$ and $l$ are positive and also satisfy the  inequalities
   \begin{eqnarray}
   r\geq p,\quad k\geq p, \quad l<2z^2r,
   \end{eqnarray}
which follow from the structure of $\varOmega$ defined by (\ref{total_Zassenhaus}) -- the power of the parameter $\tau$ equals or exceeds the power of $G({\cal B})$ and the power of $\lambda$ in all terms of the expansion. The action of each monomial of the $\varOmega$ expansion on the heat kernel $e^{-\tau{\cal A}-\lambda{\cal B}}=e^{-\lambda\tilde{\cal A}}$ under the functional trace sign generates a maximal negative power of $\lambda$ when all the derivatives are acting on the exponentiated world function. In what follows we focus on such terms dominant in $1/\lambda$, which are most troublesome from the viewpoint of the efficiency of our expansion. In virtue of (\ref{nabla_exp}) the result for these terms looks like
   \begin{eqnarray}
   \left(\frac\tau\lambda\right)^r\frac{\lambda^{\frac{k}2
   +\big(\left[\frac{l}2\right]-\frac{l}2\big)}\,
   \mathfrak{B}_k}{(M^2\lambda)^{p(z-1)}},
   \end{eqnarray}
and after the integration over $\lambda$ and $q$ in view of (\ref{q_to_tau}) it goes over into the factor
   \begin{eqnarray}
   \big(M^2\tau\big)^{(r-p)\big(1-\frac1z\big)+\frac{k}{2z}
   +\frac1z\big(\frac{l}2-\left[\frac{l}2\right]\big)}\,
   \frac{\mathfrak{B}_k}{M^k}.                 \label{factor1}
   \end{eqnarray}
This expression shows that the $\varOmega$ factor does not in fact destroy the efficiency of the curvature expansion -- in view of the inequality $r\geq p$ the growing power of the curvature is always accompanied by the growing positive power of $\tau$. Regarding the lowest powers of $\tau$ the effect of this factor in (\ref{structure}) symbolically consists in the multiplication of (\ref{factor1}) and the $v=0$ term of (\ref{A_scaling}). In this way we get the chain of lowest order powers of $\tau$ as coefficients of curvature invariants $\mathfrak{B}$
\be\begin{split}
    {\rm Tr}\,e^{-\tau{\cal D}}=&\frac{M^{\ds+1}}{(4\pi)^{\frac{\ds+1}{2}}}\,
    \frac1{\big(M^2\tau\big)^{\frac{\ds+z}{2z}}} \int \di t\,\di^\ds x\, N\gamma^{1/2}\\
    & \times  \sum\limits_{u,k,r,p,l} \big(M^2\tau\big)^{\frac{2u+k+l}{2z}+(r-p)\big(1-\frac1z\big)
    -\frac1z\,\left[\frac{l}2\right]}
    \frac{\mathfrak{B}_{2u+k}}{M^{2u+k}}+...\, \ .
\end{split}\ee
Here ellipsis denote the rest of the terms with higher powers of $\tau$, and the curvature integral $\mathfrak{B}_{2u+k}$ is the result of the composition of $\mathfrak{B}_{2u,0}$ and $\mathfrak{B}_k$, $\mathfrak{B}_{2u,0}\times\mathfrak{B}_k\to \mathfrak{B}_{2u+k}$. The new set of summation indices $n$ and $m$ (instead of $u$ and $k$),
    \begin{eqnarray}
    &&n=\frac{2u+k+l}{2}+(r-p)\,(z-1)-\left[\frac{l}2\right],\\
    &&m=2u+k,
    \end{eqnarray}
in virtue of the restrictions $u\geq 0$, $r\geq 0$, $k>0$, $0\leq p\leq r$, $p\leq k$ and $l<2z^2r$  bounds the summation range over the rest of the indices
    \begin{eqnarray}
    m\leq 2n,\quad p\leq m,\quad r\leq p+\frac{n+\left[\frac{l}2\right]-\frac{l}2}{z-1}
    \end{eqnarray}
and the expansion takes the form,
   \begin{eqnarray}
    {\rm Tr}\,e^{-\tau{\cal D}}=\frac{M^{\ds+1}}{(4\pi)^{\frac{\ds+1}{2}}}\,
    \frac1{\big(M^2\tau\big)^{\frac{\ds+z}{2z}}}
    \sum\limits_{n=0}^\infty
    \big(M^2\tau\big)^{\frac{n}{2z}}\,
    \sum\limits_{m=0}^{2n}\,
    M^{-m}{B}_{m,2n},                    \label{expansion_final_2}
    \end{eqnarray}
where ${B}_{m,2n}$ is given by a finite sum of terms
  \be
  {B}_{m,2n}\equiv  \int \di t\,\di^\ds x\, N\gamma^{1/2}\sum\limits_{r,p,l}\mathfrak{B}_{m}(r,p,l).
  \ee

The inclusion of nonzero ${\cal C}$ into $\varOmega$ does not change this property, because any new power of the negative dimensionality coefficient $\Omega_{(m)}$ is always accompanied by a positive fractional power of $\tau$. This power is composed of the original positive powers of $\tau$ and the negative powers of $\lambda$ coming from the differentiation of the exponentiated world function like in (\ref{nabla_exp}). After the integration over $\lambda$ on using (\ref{q_to_tau}) the final balance is always in favor of the positive power of $\tau$. With the example of the highest derivative part of ${\cal C}$, ${\cal C}=\Omega_{(2z-1)}\nabla^{2z-1}+...$, for the contribution of $l$-th power of ${\cal C}$ this looks like
\be
(\tau{\cal C})^l\,e^{-\sigma(X,Y)/2\lambda}\,\big|_{\,Y=X}\sim \big(\Omega_{(2z-1)}\big)^l\frac{\tau^l}{\lambda^{[(z-1/2)l]}}\sim
\big(\Omega_{(2z-1)}\big)^l\,\tau^{\frac1z\,\left[\frac{l}2\right]},
\ee
and implies that perturbation theory in $\Omega_{(m)}$ is efficient for the $\tau$-expansion of the heat kernel -- any given fractional order in $\tau$ contains a finite order polynomial in the coefficients $\Omega_{(m)}$ and their derivatives.
  
This clearly shows that the coefficient of every fractional power $n/z$ of the proper time and the power of the mass parameter is given by finite order polynomials in curvatures and background quantities $\Omega_{(m)}$ and their derivatives. This completes the proof of the efficiency of our curvature expansion and leads to (\ref{expansion_final_3})-(\ref{critBfinal}).

\subsection{Integral evaluation} \label{app:int}

Let us understand how the integrals in  \eqref{expansion1} are indeed well-defined.
The coefficients of this expansion are given by the double integrals -- the outer contour integral over $q$ and the inner integral over the parameter $\lambda$. Since the summation range over $n$ extends to $-\infty$ many of these $\lambda$-integrals are divergent at the lower limit. However, the double integral makes sense for all possible $n$ because the divergent part of the $\lambda$-integral turns out to be analytic in the vicinity of the real axes of $q$, and its integration over the contour $C$ in Fig.~\ref{fig_contour_C} encircling the positive real semi-axes gives zero. This explicit mechanism of suppression of these divergences in the resolvent method  is an advantage with respect to the  approach based on the Mellin transform \cite{D'Odorico:2015yaa}. In order to see this, we introduce a small nonzero lower limit
    \begin{eqnarray}
    \int\limits_\delta^{\pm i\infty} \di \lambda\,\lambda^{c-1} e^{\lambda q}=
    (-q)^{-c}\Gamma(c,-q\delta), \quad -q=e^{-i\pi}q,
    \end{eqnarray}
for the divergent integrals with $c\equiv n-d/2<0$. The resulting incomplete Gamma function has the following expansion of small second argument
    \begin{eqnarray}
    \Gamma(c,-q\delta)=\Gamma(c)-
    \sum\limits_{k=0}^\infty
    \frac{(-1)^k(-q\delta)^{c+k}}{k!(c+k)},   \label{incomplete}
    \end{eqnarray}
which is divergent for $\delta\to 0$ due to the first $[-c\,]+1$ terms in the infinite series in $q$. Multiplication by $(-q)^{-c}$, however, makes these terms analytic in $q$ with only integer positive powers of $q$, so that the following contour integration over $q$ leaves us only with the first term $(-q)^{-c}\Gamma(c)$ and the expansion coefficients in (\ref{expansion1}) take the form
    \begin{eqnarray}
    \frac1{2\pi i}\int\limits_C \di q\,e^{-\tau G(q)}
    \int\limits_0^{\pm i\infty}\!\di \lambda\;\,
    e^{\lambda q}\;\lambda^{n-\ds/2}=
    \frac{\Gamma(n+1-\ds/2)}{2\pi i}\int\limits_C \di q\,
    e^{-\tau G(q)}(-q)^{\ds/2-n-1}.
    \end{eqnarray}
Due to the factor $(-q)^{\ds/2-n-1}=(e^{-i\pi}q)^{\ds/2-n-1}$ the integrand is analytic in the complex plane of $q$, $0<{\rm arg}\, q<2\pi$ with the cut along the axes of real positive $q$. Since the jump on this cut or the pole is at $q=0$, this integral over the contour $C$ is nonzero. The integration can be easily performed for the function $G(q)$ of the form
    \begin{eqnarray}
    G(q)=\frac{q^z}{(M^2)^{z-1}},
    \end{eqnarray}
with an integer parameter $z$. In this case, this integral can be converted by the change of integration variable $q\mapsto v=q(M^2\tau)^{1/z}/M^2$ to the generalization of the Hankel loop integral for the Euler Gamma function
    \begin{eqnarray}
    \frac{\Gamma(b)}{2\pi i}\int\limits_C \di v\,
    (-v)^{-b}e^{-v^z}=-\frac1z\,
    \frac{\Gamma\big(\frac{1-b}z\big)}{\Gamma(1-b)},
    \quad  b=n+1-\ds/2.                                 \label{Z}
    \end{eqnarray}
This result can be obtained by deforming the contour of integration to the upper and lower shores of the cut along the positive real axes and making the contribution of the residue at the branch point $v=0$ vanishing via the integration by parts\footnote{The above derivation is not applicable for integer negative $c=-n$ corresponding to an even $d$. In this case the expansion of the incomplete Gamma function (\ref{incomplete}) should be replaced by the expression $(-q)^n\Gamma(-n,-q\delta)=-q^n\ln (-q)/n!+...$, where the only nonanalytic term was retained, which contributes to the contour $q$-integral due to the discontinuity of the logarithm at the cut of the complex $q$-plane. This leads to the same result as the right hand side of (\ref{Z}) with $c=-n$.}.

Finally,
    \begin{eqnarray}
    \frac1{2\pi i}\int\limits_C \di q\,e^{-\tau G(q)}
    \int\limits_0^{\pm i\infty}\!\di \lambda\;\,
    e^{\lambda q}\;\lambda^{n-d/2}=
    -\frac1{(4\pi)^{\ds/2}}\,
    \left(\frac{(M^2\tau)^{1/z}}{M^2}\right)^{n-d/2}
    \frac{\Gamma\big(\frac{d-2n}{2z}\big)}
    {z\Gamma\big(\frac{d-2n}{2}\big)},    \label{lambda_q_int}
    \end{eqnarray}
and substituting this result into (\ref{expansion1}) we get \eqref{expansion_final}.

%
\section{Heat kernel coefficients of the covariant Laplace operator}\label{HK_laplace}

In this appendix, we summarize the first few coefficients of the heat kernel expansion for the standard covariant Laplace operator $-\Delta$ on a $d$-dimensional manifold without boundary, which have been computed independently in \cite{DeWittbook:1965,Barvinsky:1985an,Vassilevich:2003xt},
\begin{align}
\label{eq:Lapla_as}
\Tr\left(F\; e^{-\tau (-\Delta)}\right)=\sum_{k\geq 0}  \tau^{\frac{k-d}{2}}B_{k}[\,F,-\Delta\,].
\end{align}
Their connection to the local coefficients is given in \eqref{eq:bintegrands}. The first few of them read \cite{Vassilevich:2003xt},
\begin{eqnarray}
&&B_{0}[\,F,-\Delta\,]=\frac1{(4\pi)^{d/2}}\int \di^d x\;\mathrm{tr}\,\big[\,F\,\big]. \label{a0nobou}\\
&&B_2[\,F,-\Delta\,]
         =\frac1{(4\pi)^{d/2}}\int \di^d x\; \mathrm{tr}\Big[\,\frac16\,FR\,\Big].\label{a2nobou}\\
&&   B_4[\,F,-\Delta\,]=\frac1{(4\pi)^{d/2}}\int \di^dx\; \mathrm{tr}\,\Big[F\Big(\,\frac1{30}\Delta R+\frac5{360}R^2-\frac1{180}\, R_{ij}R^{ij}\nonumber
    \\
&&\qquad\qquad
    +\frac1{180} R_{ijkl}R^{ijkl}+ \frac1{12}\,{\cal R}_{ ij}{\cal R}^{ij}\Big)\Big].\label{a4nobou}
\\
    &&B_{6}[\,F,-\Delta\,]=\frac1{(4\pi)^{d/2}}\int \di^d x\; \mathrm{tr}\Bigl[ \frac F{7!}\,\big(
    18\Delta^2 R+17\nabla_k R\nabla^k R
    -2\nabla_k R_{ij}\nabla^k R^{ij} \nonumber \\
&&\qquad\qquad -4\nabla_n R_{jk}\nabla^k R^{jn}
    +9\nabla_n R_{ijkl}\nabla^n R^{ijkl}+28R\Delta R
    -8R_{jk}\Delta R^{jk} \nonumber \\
&&\qquad\qquad
    +24R_{jk}\nabla^k \nabla_n R^{jn}
    +12R_{ijkl}\Delta R^{ijkl}
    +\frac{35}9\,R^{3}
    -\frac{14}3\,RR_{ij}R^{ij} \nonumber \\
&&\qquad\qquad
    +\frac{14}3\,R R_{ijkl}R^{ijkl}
     -\frac{208}9\,R_{jk}R^{jn}R_{n}^k
     -\frac{64}3\,R_{ij}R_{kl}R^{ik jl} \nonumber \\
&&\qquad\qquad
     -\frac{16}3\,R_{jk}R^{j}_{\;\;n l i}R^{kn l i}
     -\frac{44}9\,R_{ij kn}R^{ij l p}R^{kn}_{\;\;\;\;l p}      -\frac{80}9\,R^{ijk}_{\;\;\;\;\; n}R_{ilkp}R_{j}^{\;\;\;lnp}\big)\nonumber\\
&&\qquad\qquad
     +\frac1{360}\,F\,\big(
      8\nabla_k\Cur_{ij} \nabla^k \Cur^{ij}
     +2\nabla^j \Cur_{ij}\nabla_k\Cur^{ik}+12\Delta \Cur_{ij}\Cur^{ij}\nonumber \\
&&\qquad\qquad
     -12\Cur_{ij}\Cur^{jk}\Cur_{k}^{i}
     -6R_{ij kn}\Cur^{ij}\Cur^{kn}
     -4R_{jk}\Cur^{jn}\Cur^{k}_{n}
       +5R\Cur_{kn}\Cur_{kn}\,)\Big].\label{a6nobou}
\end{eqnarray}
where $\Cur_{ij}$ refers to the extra field strength that emerges in the presence of a spin bundle
\begin{align}
[\nabla_i,\nabla_j]\phi^A=\Cur_{ij\;\;\;B}^{\;\;\;A}\phi^B.
\end{align}

\section{Functorial properties of anisotropic operators}\label{sec:func}

In this appendix we generalize some of the standard functorial properties of covariant elliptical operators (cf. \cite{Vassilevich:2003xt,Gilkey:2004dm}) to the case of anisotropic ones.
We focus on those properties which are not already derived in the main text by other (scaling) arguments.
\begin{enumerate}
\item
The global scaling transformation of the operator \eqref{eq:dim_D} can be extended to a local transformation by allowing $\omega=\omega(X)$ to be a \emph{local} function of space and time. Following \cite{Vassilevich:2003xt}, given a general anisotropic operator ${\cal D}(D_t,\nabla_{x})$ of the form \eqref{eq:gen_D},
and transforming under the local transformation as
\be
{\cal D}\mapsto e^{-2z\omega}  {\cal D},
\ee
the infinitesimal  transformation of the heat kernel trace results in
\begin{align}
\delta_\omega \Tr\left(e^{-\tau\D}\right)=2z\tau \Tr\left(\omega \D e^{-\tau \D}\right)=-2z\tau \frac{\di}{\di\tau}\Tr\left(\omega e^{-\tau \D}\right).
\end{align}
Inserting in this expression the asymptotic expansion of the anisotropic heat kernel \eqref{eq:asymp} on both sides, we find a constraint among the the different coefficients
\be
\label{eq:ahom}
\delta_\omega \AA_k[1,\D]=(z+\ds-k) \AA_k[\omega , \D],
\ee
where the variation parameter $\omega$ plays the role of the smearing function $F$.
\item
Another useful identity can be derived by considering the class of operators $ \D-b$. If $[\D,b]=0$, then
\be
\Tr[Fe^{-\tau (\D-b)}]= \Tr[Fe^{\tau b}e^{-\tau \D}],
\ee
and
\be \label{eq:functorial_omegao}
\AA_k[F,\D-b]=\sum_{n,m}\AA_n[F, \D] \frac{b^m}{m!}, \quad n/2+z m=k/2.
\ee
The previous relation  gives information about the contribution to the effective action  of the $\Omega_0$ term in \eqref{eq:gen_D}.
In fact, by writing  $b=-\Omega_0$ and ignoring the possible appearance of derivatives of $\Omega_0$ (which is equivalent to formally assuming $[\D,\Omega_0]=0$)
the contribution of monomials of $\Omega_0$ is read from the previous equation.
\item
Finally, combining \eqref{eq:ahom} and \eqref{eq:functorial_omegao}, there is a new property that can be derived. For that, let us introduce a $\theta$ parameter and construct the following combination
\begin{align}
\D_\theta=\D-\theta b,
\end{align}
where we assume that both $\D$ and $b$ transform homogeneously under global dilatations, so that $\delta_\omega \D_\theta=e^{-2z \omega}\D_\theta$ and that $b$ is a constant. Then, using \eqref{eq:ahom} with $k=\ds+z$ we have
\begin{align}
\delta_\omega \AA_{\ds+z}[1,\D_\theta]=0.
\end{align}

Taking the variation\footnote{Defined as $\delta_{\theta}=\left.\frac{d}{d\theta}\right|_{\theta=0}$.} with respect to $\theta$ and commuting both variations
\begin{align}\label{eq:one}
\delta_\omega \delta_{\theta} \AA_{\ds+z}[1,\D_\theta]=0.
\end{align}

The coefficient $\AA_{k}[1,\D_\theta]$ is read from \eqref{eq:functorial_omegao},
and under a variation with respect to $\theta$ yields
\begin{align}\label{eq:two}
\delta_\theta \AA_{k}[1,\D_\theta]=\AA_{k-2z}[1,\D_\theta]b=\AA_{k-2z}[b,\D].
\end{align}

Finally, combining both equations \eqref{eq:one} and \eqref{eq:two}, we have
\be
\delta_\omega \AA_{\ds-z}[b,\D]=0.
\ee
for $b$ constant and transforming homogeneously under dilatations.
\end{enumerate}

\bibliography{biblio_full3}
\bibliographystyle{JHEP}
\end{document}